\documentclass[aps,prc,letterpaper,11pt,twoside,tightenlines,nofootinbib,showpacs,preprint,superscriptaddress]{revtex4}
\usepackage{amsmath}
\usepackage{graphicx}
\usepackage{subfigure}
\usepackage{color}
\RequirePackage[colorlinks,citecolor=blue,urlcolor=blue,linkcolor=blue]{hyperref}

\newcommand{\mev}{\textrm{ MeV}}
\newcommand{\gev}{\textrm{ GeV}} 

\newcommand{\nn}{\nonumber}

\begin{document}
 
\title{Investigation of {\boldmath$J/\psi \to  \gamma\, \pi^0 \eta (\pi^+\pi^-, \pi^0\pi^0)$} radiative decays including final-state
  interactions}

\author{C. W. Xiao}
\affiliation{School of Physics and Electronics, Central South University, Changsha 410083, China}
\affiliation{Institut  f\"{u}r Kernphysik (Theorie), Institute for Advanced Simulation, and J\"ulich Center for Hadron Physics,
  Forschungszentrum J\"ulich, D-52425 J\"{u}lich, Germany}

\author{U.-G.~Mei{\ss}ner}
\affiliation{Helmholtz-Institut f\"ur Strahlen- und Kernphysik, and Bethe Center for Theoretical Physics, Universit\"at Bonn, D-53115  Bonn, Germany}
\affiliation{Institut  f\"{u}r Kernphysik (Theorie), Institute for Advanced Simulation, and J\"ulich Center for Hadron Physics,
  Forschungszentrum J\"ulich, D-52425 J\"{u}lich, Germany}
\affiliation{Ivane Javakhishvili Tbilisi State University, 0186 Tbilisi, Georgia}

\author{J. A. Oller}
\affiliation{Departamento de F\'isica, Universidad de Murcia, E-30071 Murcia, Spain}

\begin{abstract}

  We revisit the coupled channel $K\bar{K}$ interactions and dynamically generate the resonances $f_0(980)$ and $a_0(980)$ within both the isospin and 
  the physical bases. The $f_0(980)-a_0(980)$ mixing effects are  generated in the scattering amplitudes of the coupled channels with the 
  physical basis,  which exploits the important role of the $K\bar{K}$ channel in the dynamical nature of these resonances. With the scattering 
  amplitudes obtained, we investigate the $f_0(980)$ and $a_0(980)$ contributions to the $J/\psi\to \gamma\eta\pi^0$, $J/\psi\to \gamma\pi^+\pi^-$ 
  and  $J/\psi\to \gamma\pi^0\pi^0$ radiative decays through the final-state interactions. We obtain the corresponding branching fractions 
  $Br(J/\psi\to \gamma a_0(980) \to  \gamma\eta\pi^0) = (0.47\pm0.05) \times 10^{-7}$, $Br(J/\psi\to \gamma f_0(980) \to  \gamma\pi^+\pi^-) =
  0.37 \times 10^{-7} - 1.98 \times 10^{-6}$, $Br(J/\psi\to \gamma f_0(980) \to  \gamma\pi^0\pi^0) = 0.18 \times 10^{-7} - 9.92 \times 10^{-7}$, 
  and predict $Br(J/\psi\to \gamma a_0(980)) = 1.72 \times 10^{-8} - 3.07\times 10^{-7}$ and $Br(J/\psi\to \gamma f_0(980)) = 1.86 \times 10^{-8} 
  - 1.89\times 10^{-5}$. These fractions are within the upper limits of the experimental measurements.
  
\end{abstract}

\pacs{}
\maketitle
\date{}

\section{Introduction}

The new pentaquark candidates, i.e. the so-called $P_c^+$ states, found by LHCb collaboration \cite{Aaij:2015tga,Aaij:2019vzc}
have caught much attention
from both theorists and experimentalists to understand the properties of the ``exotic'' states in the QCD spectrum, see the reviews
\cite{Chen:2016qju,Hosaka:2016pey,Chen:2016spr,Lebed:2016hpi,Esposito:2016noz,Guo:2017jvc,Ali:2017jda,Olsen:2017bmm,Karliner:2017qhf,Yuan:2018inv}.
To understand the properties of these ``exotic'' candidates is one of the main tasks in contemporary particle physics.
This is not a trivial issue because one is dealing with non-perturbative strong interactions.
Similar problems have been faced since long in the light quark sector.
The two famous states, $f_0(980)$ and $a_0(980)$, found around the end of the 1960s \cite{Astier:1967zz,Ammar:1969vy,Defoix:1969qx,Protopopescu:1973sh,Hyams:1973zf}, still require more investigations to accurately determine their properties and nature. In the literature, they are  assigned as
$q \bar{q}$ states \cite{Godfrey:1985xj,Morgan:1990kw,Morgan:1993td}, $qq \bar{q}\bar{q}$ states \cite{Jaffe:1976ig,Jaffe:1976ih,Achasov:1980tb},
scalar glueballs  \cite{Achasov:1980tb,Au:1986mq}, $K \bar{K}$ cusp effects \cite{Flatte:1976xu}, or,
meson-meson states 
\cite{Weinstein:1982gc,Weinstein:1990gu,Zou:1994ea,Janssen:1994wn,Tornqvist:1995ay,Oller:1997ti,Locher:1997gr,Oller:1998hw,Oller:1998zr}. In this
regard, Ref.~\cite{Baru:2003qq} concluded that the $f_0(980)$ and $a_0(980)$ are not elementary states based on a Flatt\'e parameterization analysis
around the $K\bar{K}$ threshold and rather general considerations on quantifying the compositeness of a near-threshold resonance.
This was later further elaborated in Refs.~\cite{Baru:2004xg,Baru:2010ww}. Indeed, within the Flatt\'e model, the $a_0(980)$ always appears 
as a $K \bar{K}$ cusp
\cite{Flatte:1976xu,Baru:2003qq,Baru:2004xg,Bugg:1994mg}. The cusp-like structure of the $a_0(980)$ is also seen from a first-principle
lattice calculation~\cite{Dudek:2016cru,Guo:2016zep}.
Based on a dispersive analysis of the experimental data, the pole of the $f_0(980)$
was precisely determined within a  model-independent approach based on a set of Roy-like equations in
Ref.~\cite{GarciaMartin:2011jx}. 
Also based on the use of the Roy equations for the  isoscalar $\pi\pi$ S-wave \cite{Colangelo:2001df}, 
  the Refs.~\cite{GarciaMartin:2011jx,Caprini:2005zr} determined the mass and width of the $f_0(500)$ resonance (also called $\sigma$)
\cite{pdg2018}. 
Other detailed determinations of the properties of the scalar resonances, and in particular of the $f_0(500)$, were undertaken
in Refs.~\cite{albaladejo.190504.1,albaladejo.190504.2,guo.190504.1}.

On the other hand,  Ref.~\cite{Oller:1997ti} finds  poles corresponding to the $f_0(500)$, $f_0(980)$ and $a_0(980)$ resonances.
This study has only one free parameter in the unitarity loop functions and
employs the lowest order chiral Lagrangian to provide the potential used in a
Bethe-Salpeter equation. 
It was further noticed in this
reference that, within the approach followed, the $f_0(980)$ stems from the $K\bar{K}$ dynamics, as a bound state in the decoupling limit with
$\pi\pi$, while the $a_0(980)$ disappears when the two channels $K\bar{K}$ and $\pi\eta$ are decoupled. This conclusion is also in agreement
with the earlier results of Ref.~\cite{Janssen:1994wn} in a meson-exchange model.
The line shape of the $a_0(980)$ was also studied in Ref.~\cite{Oller:1998hw}, which concludes that the $a_0(980)$ looks like a cusp behavior,
a result further confirmed by the analyses in Refs.~\cite{Oller:1998zr,Guo:2016zep,GomezNicola:2001as,Guo:2011pa}.
Note that, from their results, one can conclude that the $K^+K^-$ states play a prominent role in the generation through 
S-wave meson-meson interactions of the $f_0(980)$ and $a_0(980)$ resonances.
The interested reader can also see the review of Ref.~\cite{Guo:2017jvc} for other discussions concerning  molecular states.

Given the nature of the $f_0(980)$ and $a_0(980)$ as dynamically generated resonances, which have similar masses in the nearby of the
$K\bar{K}$ threshold, we consider their mixing through the unitarity $K\bar{K}$ loop
because of the difference in the neutral and charged kaon masses. 
This mechanism is analogous to the one driving to a sizeable
mixing of the  $\rho(770)-\omega(782)$ resonances, as pointed out in Ref.~\cite{Achasov:1979xc} already a long time ago.
Further investigations on the $f_0(980)-a_0(980)$ mixing mechanisms can be found in
Refs.~\cite{Kerbikov:2000pu,Close:2000ah,Grishina:2001zj,Close:2001ay,Achasov:2002hg,Kudryavtsev:2002uu,Achasov:2003se,Achasov:2004ur,Wu:2007jh,Hanhart:2007bd,Wu:2008hx}, where various reactions are proposed to detect the mixing signals experimentally.
Inspired by the theoretical results \cite{Wu:2007jh,Hanhart:2007bd,Wu:2008hx}, the mixing effect was firstly reported by the BESIII collaboration
\cite{Ablikim:2010aa} in the processes $J/\psi \to \phi f_0(980) \to \phi a_0^0(980) \to \phi\pi^0\eta$ and
$\chi_{c1} \to \pi^0 a_0^0(980) \to \pi^0 f_0(980)\to \pi^0\pi^+\pi^-$.
Updated results with higher statistical significance were recently given recently in \cite{Ablikim:2018pik}.
These reactions have been  analyzed by Refs.~\cite{Roca:2012cv,Bayar:2017pzq}
within the so-called chiral unitary approach (ChUA) \cite{Oller:1997ti,Oller:1998hw,Oset:1997it,Oller:2000fj},
and by Ref.~\cite{Sekihara:2014qxa} which employs a  Flatt\'e  parameterization.

The branching ratio of the $\phi\to K^0 \bar{K}^0 \gamma$
radiative decay was obtained in Ref.~\cite{Nussinov:1989gs} with the $f_0(980)$ resonance contribution stemming from
a triangle $K\bar{K}$ loop.
Along similar lines, the $\phi$ meson radiative decays were investigated in
detail in  Refs.~\cite{Achasov:1987ts} with a dispersive approach, where the branching ratios of $\phi\to \pi^0 \eta \gamma$, $\phi\to \pi\pi\gamma$,
$\phi\to \gamma K^+K^-$ and $\phi\to \gamma K^0\bar{K}^0$ were given.
These $\phi$ meson radiative decays results were also considered by many other references
\cite{LucioMartinez:1990uw,Close:1992ay,Bramon:2000vu,Bramon:2001un,Close:2001ay,Bramon:2002iw}, with the triangle $K\bar{K}$ loop playing an essential role.
Further, Refs.~\cite{Oller:1998ia,Marco:1999df,Palomar:2003rb} introduced the use of the ChUA.

By the replacement of the $s\bar{s}$ in the $\phi(1020)$  by a $c\bar{c}$ in the $J/\psi$, one is naturally driven to consider
the $J/\psi$ radiative decays. 
In this regard, the BESIII Collaboration has observed recently the  radiative decay $J/\psi\to \gamma\eta\pi^0$ and obtained
its branching ratio \cite{Ablikim:2016exh}.
A main goal of our present work is to study several $J/\psi$ radiative decay channels within ChUA, where we mainly investigate the resonance
contributions of the $a_0(980)$ and $f_0(980)$ and contributions from their mixing.

In this work, we firstly revisit  in Sec.~\ref{sec.190501.1} the coupled-channel $K\bar{K}$ interactions
by using the ChUA using the  isospin and charged channels.
In the following section we investigate the $J/\psi\to \gamma\eta\pi^0$ radiative decay 
to check the resonance contributions of the $a_0(980)$.
As a by product, the decays $J/\psi\to \gamma\pi^+\pi^-$ and  $J/\psi\to \gamma\pi^0\pi^0$
are also considered and the  $f_0(980)-a_0(980)$ mixing effects are discussed.
Finally,  our conclusions are given in Sec.~\ref{sec.190501.3}.

\section{The $K\bar{K}$ interactions revisited}
\label{sec.190501.1}

Within the ChUA, the scattering amplitudes are calculated by solving the coupled channel Bethe-Salpeter equation
with the on-shell factorization \cite{Oller:1997ti,Oller:1998hw,Oset:1997it,Oller:2000fj},
\begin{equation}
T = [1 - V \, G]^{-1}\, V,
\label{eq:BS}
\end{equation}
where $G$ is a diagonal matrix with the unitarity loop functions of the intermediate states
and  $V$ is the matrix that contains the potentials evaluated from the lowest order chiral Lagrangian.\footnote{Given  the exploratory 
aim of our study on the mixing between the $f_0(980)$ and $a_0(980)$ resonances in
$J/\psi$ radiative decays, and since the experimental data can be well reproduced by unitarizing the leading order
$V$ matrix (as shown below), we do not take into account in this work the contributions from the next-to-leading-order chiral perturbation
theory amplitudes. The latter were already applied in the study of the scalar meson-meson scattering and its spectroscopy in several works
in the literature, e.g. in
\cite{Oller:1998hw,Guo:2016zep,Guo:2011pa,albaladejo.190504.2,guo.190504.1,Pelaez:2006nj,Pelaez:2010fj}.}
The loop functions in $G$ are regularized either by employing a three-momentum cutoff $q_{\max}$ \cite{Oller:1997ti,Oller:1998hw,Oset:1997it} 
or in terms of a subtraction constant $a(\mu)$, within the dimensional-regularization
scheme introduced in Ref.~\cite{Oller:2000fj}. These are the only free parameters in the calculation of $T(s)$.

By using a three-momentum cutoff, we have
\begin{equation}
G_{ii} (s) = \int_0^{q_{\rm max}} \frac{d^3 \vec{q}}{(2\pi)^{3}}\frac{\omega_1+\omega_2}{2\omega_1\omega_2}\,\frac{1}{s-(\omega_1+\omega_2)^2+i\varepsilon},
\label{eq:Gco}
\end{equation}
where $\omega_i = \sqrt{{\vec{q}_i\!}\,^2+m_i^2},~(i =1,\,2, )$. 
It turns out by the fit to data discussed below, that $q_{\rm max}\simeq 900\mev$ \cite{Aceti:2015zva},
which is a natural value \cite{Oller:2000fj} for the short-distance scale of the strong interactions in such hadronic processes.
This value for $q_{\rm max}$  is equivalent to $\Lambda=\sqrt{m_K^2+q_{\rm max}^2}\simeq 1.03 \gev$, that corresponds to the maximum energy 
of the intermediate $K$ meson as introduced in Ref. \cite{Oller:1997ti}.

In our case, we revisit the $K\bar{K}$ interactions in coupled channels following Ref.~\cite{Oller:1997ti}.
The elements of the matrix $V$  in the isospin basis are
\begin{eqnarray}
  \label{190501.1}
V^{I=0}_{11} (s) &=& -\frac{1}{2f_\pi^2}(2s - M_\pi^2), \;
V^{I=0}_{12} (s) = -\frac{\sqrt{3}}{4f_\pi^2}s, \;
V^{I=0}_{22} (s) = -\frac{3}{4f_\pi^2}s,\\
  \label{190501.1b}
V^{I=1}_{11} (s) &=& -\frac{1}{3f_\pi^2} M_\pi^2, \;
V^{I=1}_{12} (s) = \frac{\sqrt{6}}{36f_\pi^2}(9s -8M_K^2 - M_\pi^2-3M_\eta^2), \;
V^{I=1}_{22} (s) = -\frac{1}{4f_\pi^2}s,
\end{eqnarray}
with the pion decay constant $f_\pi=93 \mev$ \footnote{We take this value for comparison of our
  results with Refs. \cite{Oller:1997ti,Oller:1998hw}. Now its updated value is 92.1~MeV \cite{pdg2018}, which affects 
  only slightly the fitted value of  $q_{\rm max}$ and does not change  our final results.}. 
In the isoscalar scattering, the channels $\pi\pi$ and $K\bar{K}$ are represented by the labels 1 and 2; 
for $I=1$, the channel $\pi^0 \eta$ is denoted by 1 and the heavier $K \bar K$ one by 2.
The matrix elements  in the isospin basis given in Eqs.~\eqref{190501.1} and \eqref{190501.1b} are deduced 
from the ones in the charge  (physical) basis. The latter are 
calculated from the lowest order chiral Lagrangian and are collected in Table~\ref{tab:vmatr}.
In addition, one should also do the S-wave projection of these amplitudes, for more details on the formalism used
we refer to  Ref.~\cite{Oller:1997ti}.

\begin{table}[ht]
     \renewcommand{\arraystretch}{1.5}
     \setlength{\tabcolsep}{0.4cm}
\centering
\caption{The matrix elements $V_{ij}$ of the potential between states in the charge basis in the isospin limit.}
\label{tab:vmatr}
\begin{tabular}{|l|l|}
\hline
\ Channel & \ potential \\
\hline
$K^+K^- \rightarrow K^+K^-$  & $-{1\over3f_\pi^2}(s+t-2u+2M_K^2)$ \\
\qquad\quad\;\,$\rightarrow K^0\bar K^0$ & $-{1\over6f_\pi^2}(s+t-2u+2M_K^2)$ \\
\qquad\quad\;\,$\rightarrow \pi^+\pi^-$ & $-{1\over6f_\pi^2}(s+t-2u+M_K^2+M_\pi^2)$ \\
\qquad\quad\;\,$\rightarrow \pi^0\pi^0$ & $-{1\over12f_\pi^2}(2s-t-u+2M_K^2+2M_\pi^2)$ \\
\qquad\quad\;\,$\rightarrow \pi^0\eta$ & $-{1\over12\sqrt{3}f_\pi^2}[3(2s-t-u)-2M_K^2+2M_\pi^2]$ \\
\hline
$K^0\bar K^0\rightarrow K^0\bar K^0$  & $-{1\over3f_\pi^2}(s+t-2u+2M_K^2)$ \\
\qquad\quad$\rightarrow \pi^+\pi^-$ & $-{1\over6f_\pi^2}(s-2t+u+M_K^2+M_\pi^2)$ \\
\qquad\quad$\rightarrow \pi^0\pi^0$ & $-{1\over12f_\pi^2}(2s-t-u+2M_K^2+2M_\pi^2)$ \\
\qquad\quad$\rightarrow \pi^0\eta$ & $-{1\over12\sqrt{3}f_\pi^2}[-3(2s-t-u)+2M_K^2-2M_\pi^2]$ \\
\hline
$\pi^+\pi^-\rightarrow \pi^+\pi^-$  & $-{1\over3f_\pi^2}(s+t-2u+2M_\pi^2)$ \\
\qquad\quad$\rightarrow \pi^0\pi^0$ & $-{1\over3f_\pi^2}(2s-t-u+M_\pi^2)$ \\
\qquad\quad$\rightarrow \pi^0\eta$ & ---  \\
\hline
$\pi^0\pi^0\rightarrow \pi^0\pi^0$  & $-{1\over f_\pi^2}M_\pi^2$ \\
\qquad\;\,$\rightarrow \pi^0\eta$  & ---  \\
\hline
$\pi^0\eta\rightarrow \pi^0\eta$ & $-{1\over3f_\pi^2}M_\pi^2$ \\
\hline
\end{tabular}
\end{table}

\begin{figure}
\centering
\begin{tabular}{ll}
\includegraphics[scale=0.7]{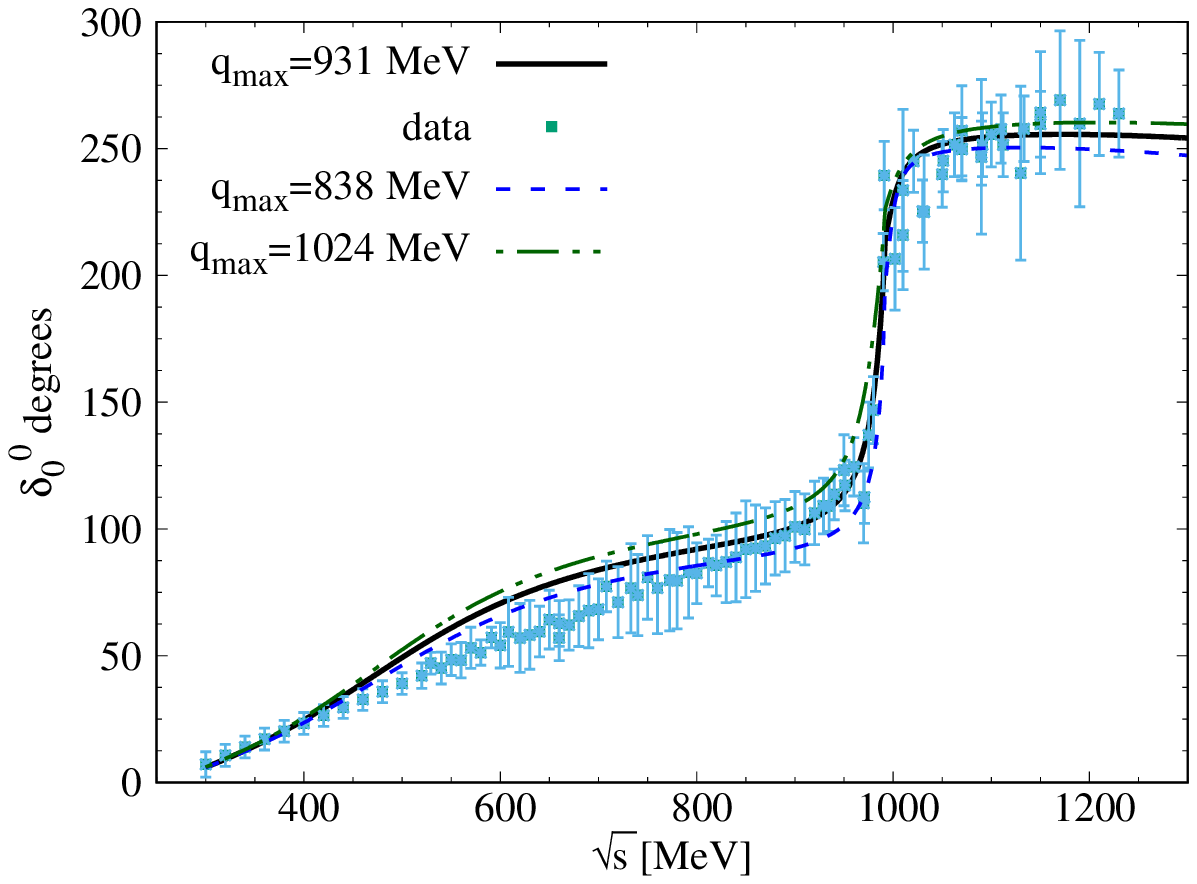} &
\includegraphics[scale=0.7]{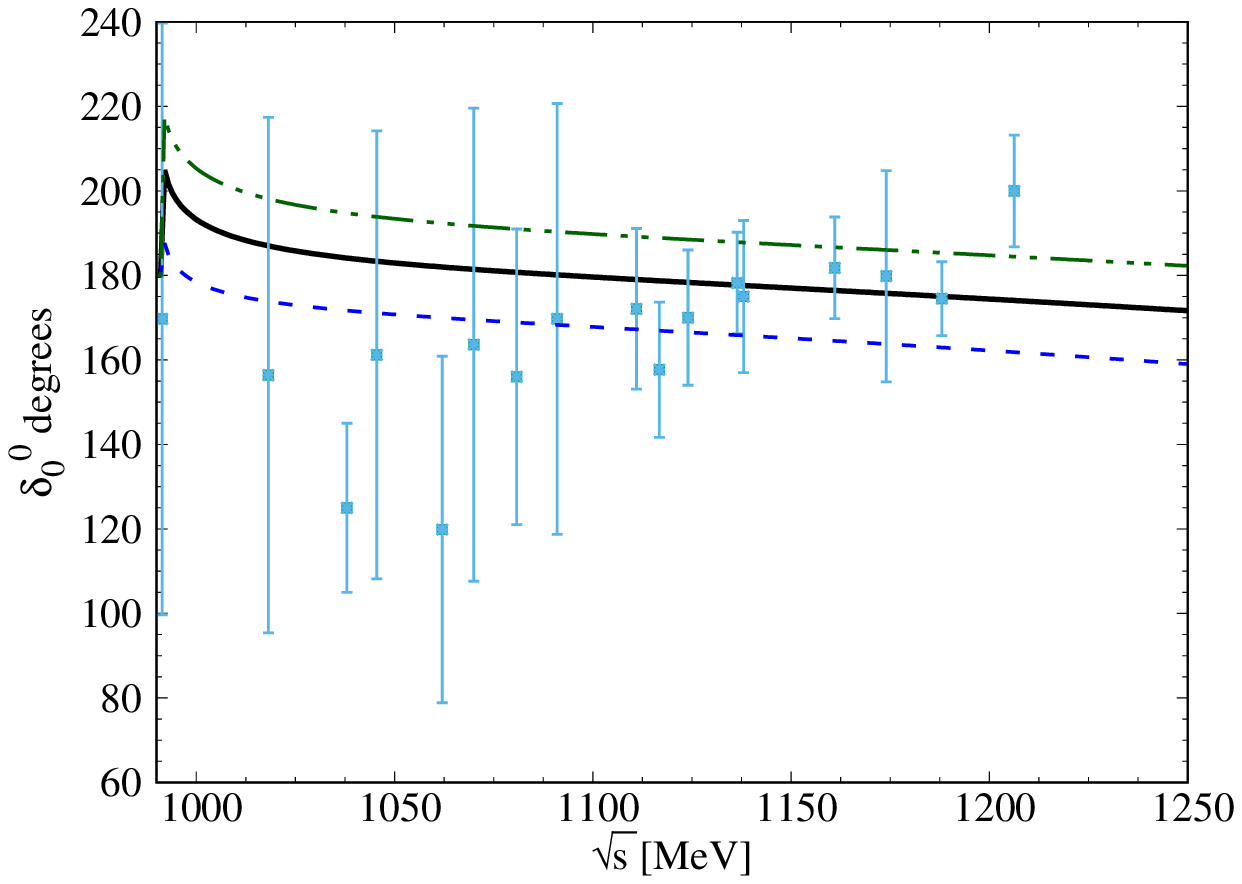} \\
\includegraphics[scale=0.7]{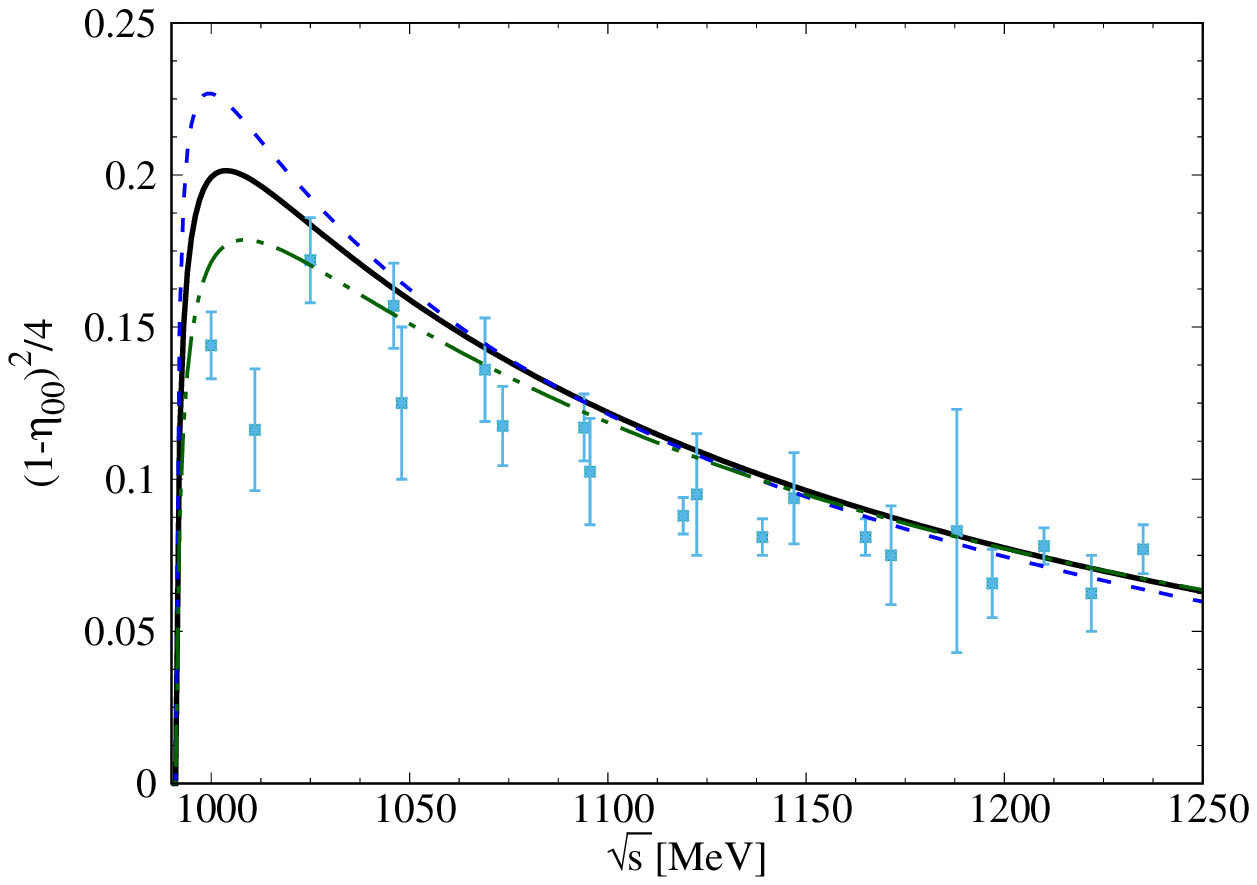}
\end{tabular}
\caption{From left to right and top to bottom:
 Results from the fit to the isoscalar scalar $\pi\pi$  phase shifts, $\pi\pi\to K\bar{K}$ phase shits,  
and the inelasticity $(1-\eta_{00}^2)/{4}$, with $\eta_{00}$ the elasticity parameter.
In the plots, the results of the best fit are indicated by the (black) solid lines and
the other lines use values of the cutoff that differ by $\pm 10\%$ from its central value.
These curves and the corresponding values of the cutoff are indicated inside the first plot. 
The references to the experimental data can be found in \cite{Oller:1998zr}.}
\label{fig:fitres1}
\end{figure}

\begin{figure}
\centering
\begin{tabular}{ll}
\includegraphics[scale=0.7]{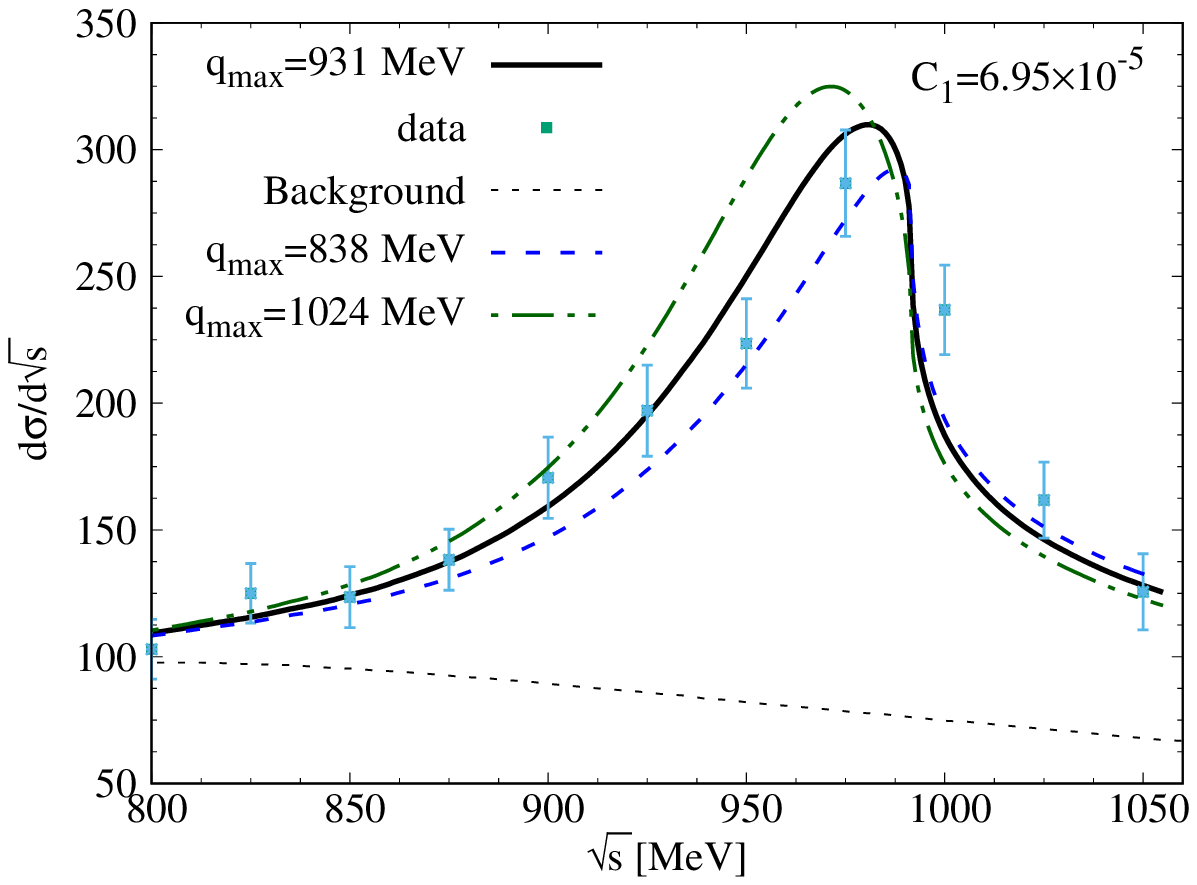} & 
\includegraphics[scale=0.7]{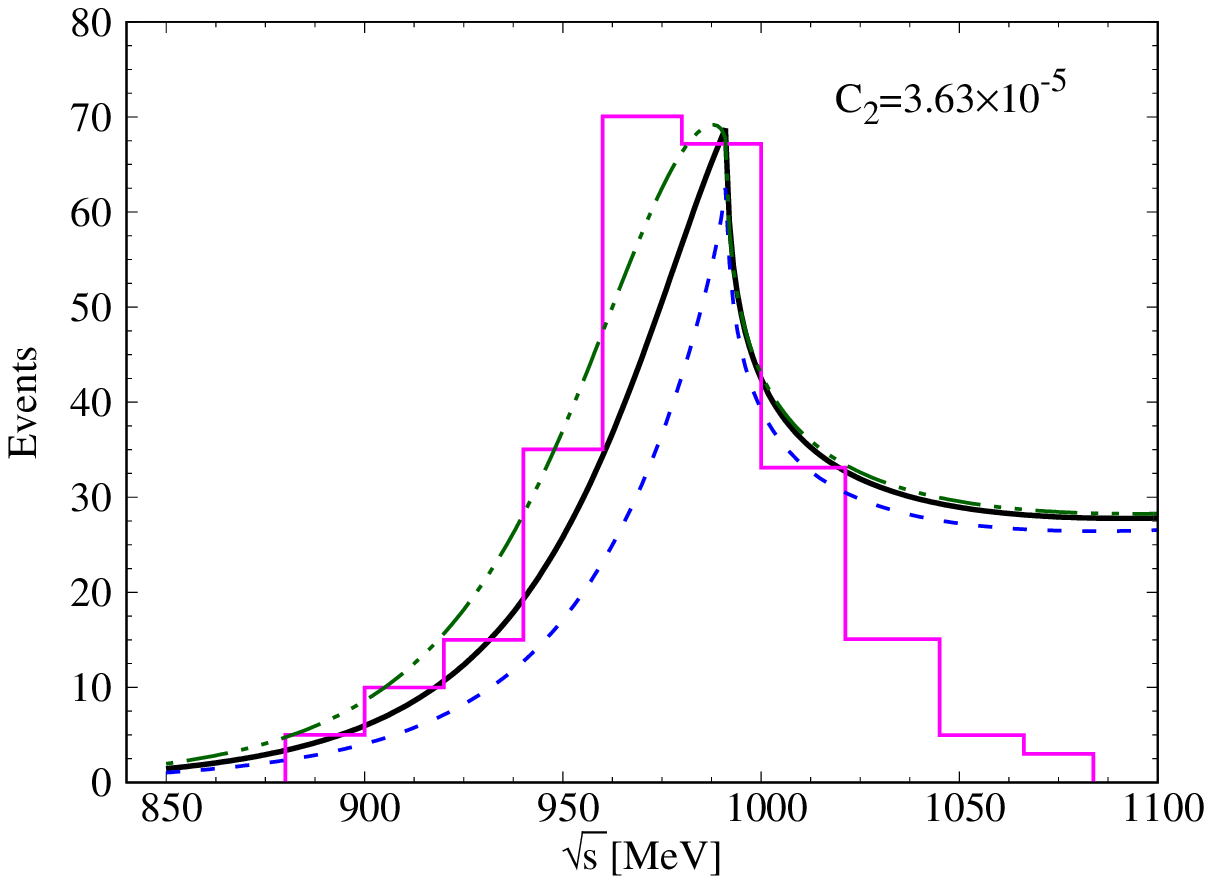}
\end{tabular}
\caption{Fit results for
the $\pi\eta$ mass distributions of the reactions $pp\to p_f(\pi^+\pi^-\eta)p_s$ (left) \cite{hp.190506.1} and 
$K^-p\to \Sigma^+(1385)\pi^-\eta$ (right) \cite{hp.190506.2}. 
For the former reaction the background is indicated by the (black) dashed line running at the bottom. 
The meaning of the rest of the lines is the same as in Fig.~\ref{fig:fitres1}.}
\label{fig:fitres2}
\end{figure}

For the only one free parameter ($q_{\rm max}$), the value  $q_{\rm max}\simeq 900\mev$ is used in Ref. \cite{Aceti:2015zva}. 
In order to estimate the uncertainty in the value of $q_{\rm max}$ (and its impact in our results),
we perform now a combined fit of the experimental data used in Refs. \cite{Oller:1997ti,Oller:1998hw,Oller:1998zr}.
These data are shown in Figs.~\ref{fig:fitres1} and \ref{fig:fitres2}. They comprise, on the one hand, 
the isoscalar scalar $\pi\pi$ elastic phase shifts,   the $\pi\pi\to K\bar{K}$ ones, 
and the inelasticity  $(1-\eta_{00}^2)/{4}$, with $\eta_{00}$ the isoscalar scalar elasticity parameter.
On the other hand, we also show in Fig.~\ref{fig:fitres2} two $\pi\eta$ event distribution sensitive to the modulus squared of 
the isovector scalar $\pi\eta$ elastic partial-wave amplitude (PWA). For their calculation we employ the same formulas as
in Refs.~\cite{Oller:1998zr,Oller:1997ti}, which read
\begin{align}
\frac{dN}{dE}&=C_1q_{\pi\eta}|t_{\pi\eta\to K\bar{K}}|^2+\alpha+\beta E~,\\
\frac{d\sigma_{\pi \eta}}{dE}&=C_2 q_{\pi\eta} |t_{\pi\eta\to \pi\eta}|^2~,  
\end{align}
where $\alpha$ and $\beta$ reproduce the incoherent background and their values are taken from the original Ref.~\cite{hp.190506.2}.
The first line is used in the left plot of  Fig.~\ref{fig:fitres2} and the last one for its right plot.

The fits obtained are shown in Figs.~\ref{fig:fitres1} and \ref{fig:fitres2}.
There. the (black) solid lines correspond to the best-fit results with the cutoff 
$q_{\rm max}= (931 \pm 60) \mev$ 
and the normalization constants of the mass distribution are
$C_1=(7.0\pm 6.5) \times 10^{-5}$~MeV$^{-2}$ 
and $C_2=(3.6\pm 1.0) \times 10^{-5}$~$\mu b/{\rm GeV}/{\rm MeV}$. 
The error-bars have been enlarged so as to take into account that the  $\chi^2_{dof}>1$. 
To be conservative we finally take an uncertainty of a 10\%  in the three-momentum cutoff, and the resulting
curves with values of the cutoff a 10\% larger and smaller than the best-fit value are also shown
in Figs.~\ref{fig:fitres1} and \ref{fig:fitres2}.

\begin{figure}
\centering
\includegraphics[scale=0.6]{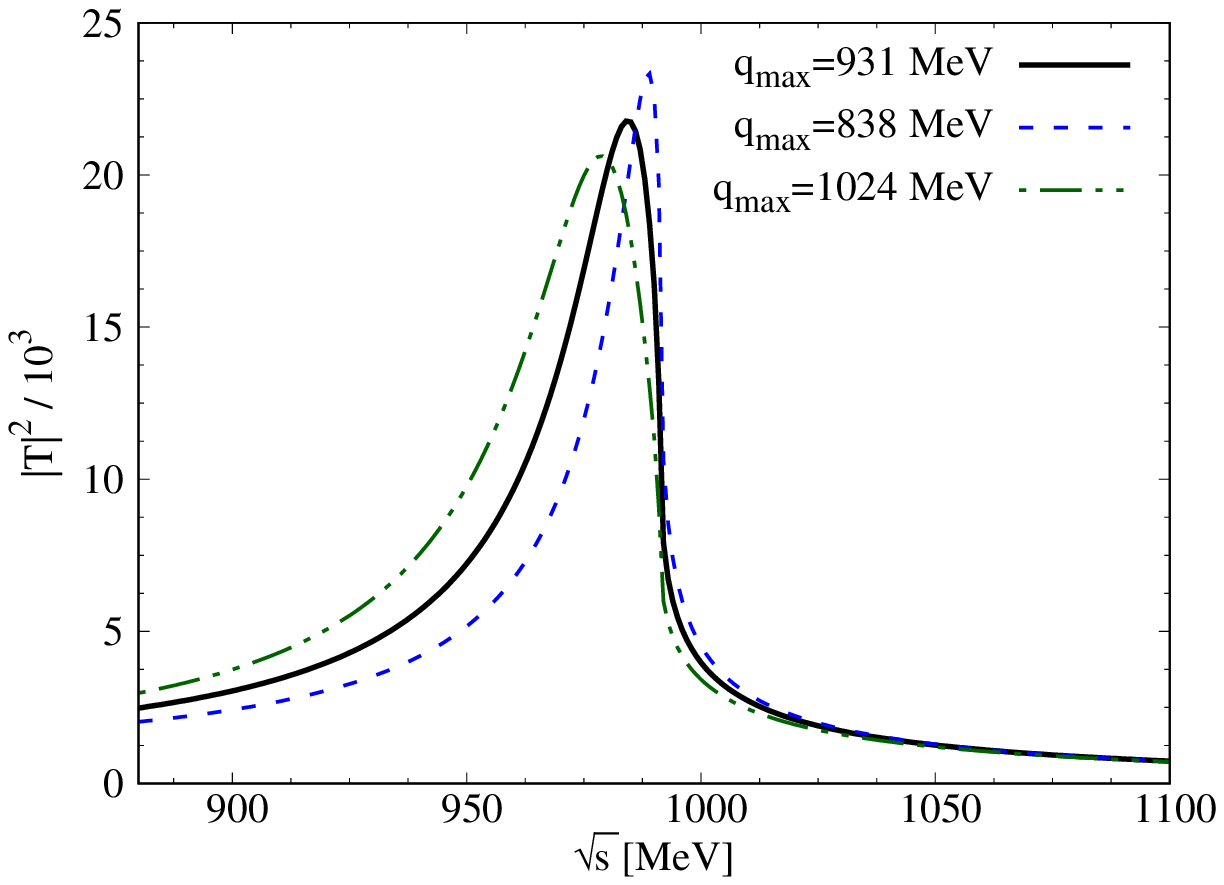}
\includegraphics[scale=0.6]{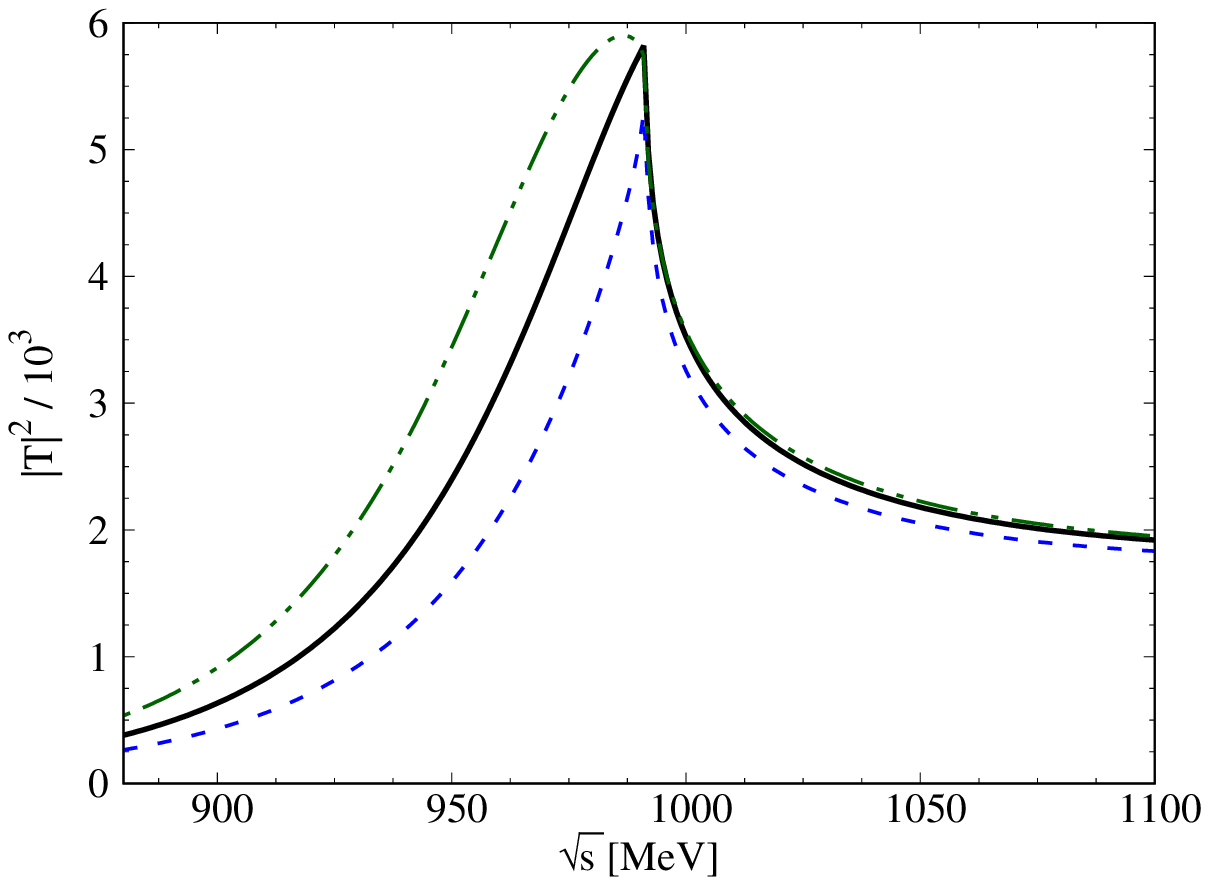}
\caption{The modulus squared of several amplitudes in the isospin basis is plotted.  
The left panel corresponds to $|T_{12}^{I=0}|^2$ [$f_0(980)$], and the right one to 
$|T_{22}^{I=1}|^2$ [$a_0(980)$]. The meaning of the lines is the same as in Fig.~\ref{fig:fitres1}.}
\label{fig:tsqf0a0}
\end{figure}

\begin{figure}
\centering
\includegraphics[scale=0.8]{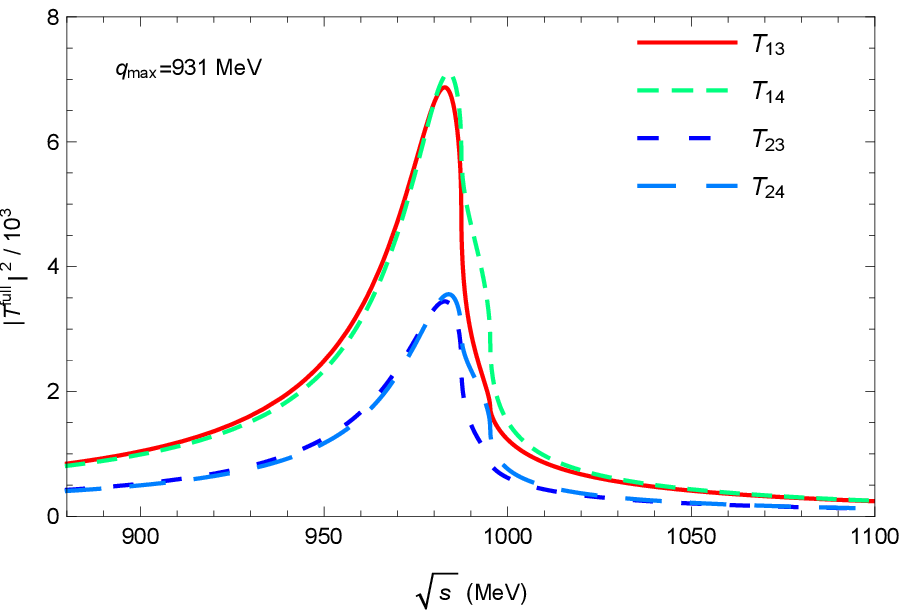}
\includegraphics[scale=0.8]{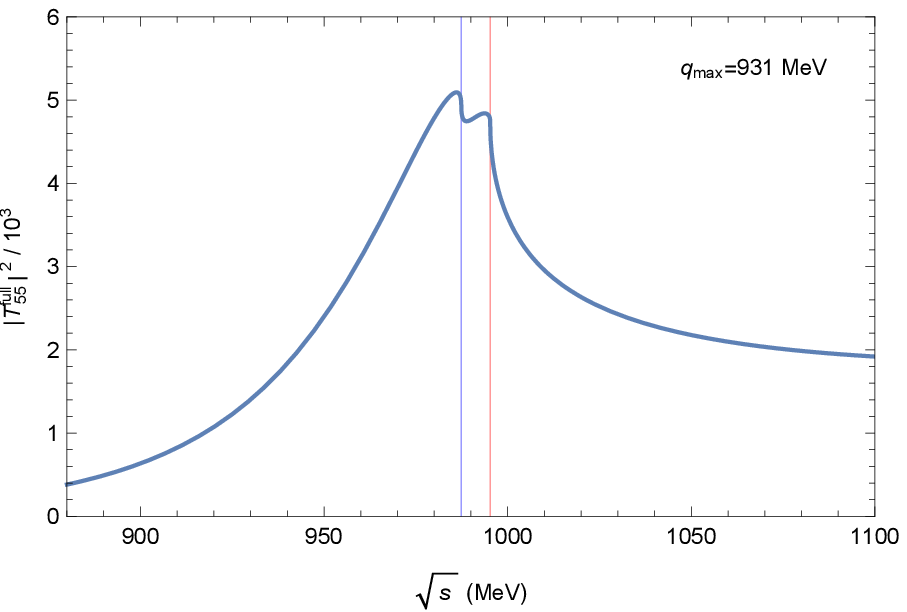}
\caption{The modulus squared of several amplitudes $T_{ij}^{full}(s)$ in the charge basis around the
    $f_0(980)$ resonance is plotted in the left panel. 
    The quantity $|T_{55}^{full}|^2$ is drawn in the right panel around the  $a_0(980)$ resonance.
}
\label{fig:tsqf0a02}
\end{figure}

\begin{figure}
\centering
\includegraphics[scale=1.0]{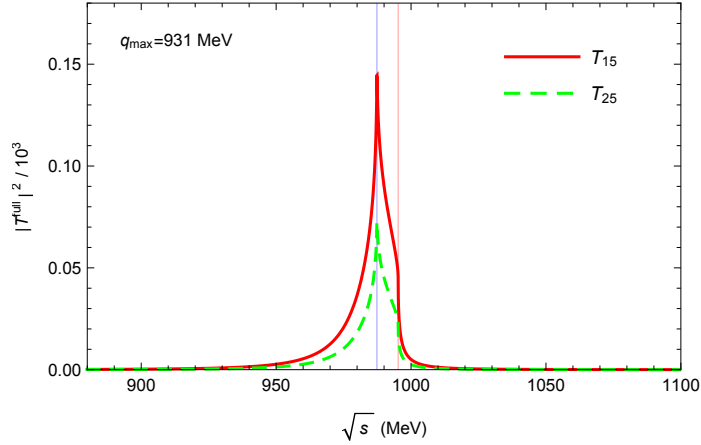}
\caption{The dynamically generated $f_0(980)-a_0(980)$ mixing effects in the modulus squared of the amplitudes $|T_{15}|^2$ and $|T_{25}|^2$.
The vertical lines lie at the energies of the  $K^+K^-$ and $K^0\bar{K}^0$ thresholds, from lighter to heavier, respectively.}
\label{fig:tsqf0a03}
\end{figure}

\begin{figure}
\centering
\includegraphics[scale=0.6]{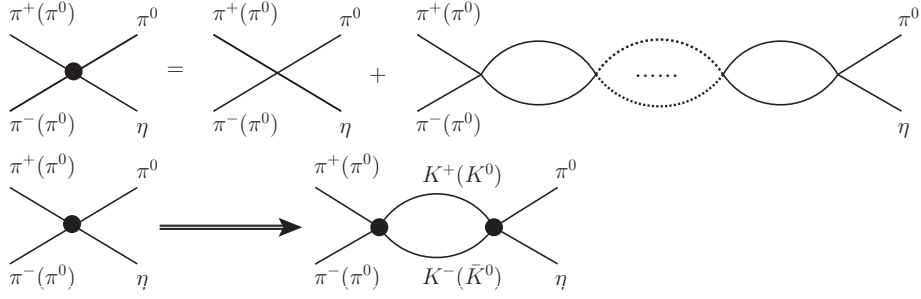}
\caption{Graphical schematic representation of the $f_0(980)-a_0(980)$ mixing effects in the amplitudes $T_{15}$ and $T_{25}$, where the
  filled black circles correspond to the full scattering amplitudes. In the first row the iteration of the potentials is represented
  by the chain of unitarity loops indicated by the ellipsis.}
\label{fig:diag}
\end{figure}

The resulting moduli squared of the partial-wave amplitudes in the isospin basis
are shown in Fig.~\ref{fig:tsqf0a0}, which are consistent with the results of Refs.~\cite{Oller:1997ti,Oller:1998hw}.
In the charge basis,  where the indices 1 to 5 denote the channels of $\pi^+\pi^-$, $\pi^0\pi^0$, $K^+K^-$, $K^0\bar{K}^0$ and $\pi^0\eta$, 
respectively,  we obtain the results shown in Fig. \ref{fig:tsqf0a02}.  
 The resonance signals corresponding to the $f_0(980)$ and $a_0(980)$ are also generated and can be seen clearly in these figures.
Since the $a_0(980)$ is strongly affected by the $K\bar{K}$ threshold, we also display in
the right panel of Fig.~\ref{fig:tsqf0a02} the energy gap of about $8\mev$ between the charged and neutral $K\bar{K}$ thresholds where the two peaks lie.
These two thresholds are indicated in this figure by the two vertical lines, with the threshold of the neutral kaons at the higher energy.
We have checked that the results of Figs.~\ref{fig:tsqf0a0} and \ref{fig:tsqf0a02} are consistent with each other, once the isospin and 
the charge bases are used, respectively, as discussed e.g. in Ref. \cite{Oset:1997it}.
Even though the matrix elements for the potential 
are zero among the channels  $\pi^+\pi^-$ ($\pi^0\pi^0$) and $\pi^0\eta$  in the charge basis, 
since they are isospin violating transitions, cf. Table. \ref{tab:vmatr}, 
nonzero scattering amplitudes result  because of the coupled-channel dynamics,  
 see Fig. \ref{fig:tsqf0a03}. 
(Similar results are also pointed out in Ref. \cite{Bayar:2017pzq}).
Interestingly, we can clearly see in this figure the $f_0(980)-a_0(980)$ mixing
in the scattering amplitudes $|T_{15}|^2$ and $|T_{25}|^2$,
as predicted in Ref. \cite{Achasov:1979xc}.
These results can be easily understood, 
since the resonances $f_0(980)$ and $a_0(980)$ are dynamically generated by using the ChUA
with coupled channels, as shown in Fig. \ref{fig:tsqf0a02}.
Thus, the amplitudes $|T_{15}|^2$ and $|T_{25}|^2$ contain the mixing effects of the dynamical diagram of Fig.~1
in Ref.~\cite{Achasov:1979xc}, as illustrated in Fig.~\ref{fig:diag},
because of the resummation series inherent to Eq.~\eqref{eq:BS}, as represented schematically in the first line of
  Fig.~\ref{fig:diag}.

\begin{figure}
\centering
\includegraphics[scale=1.0]{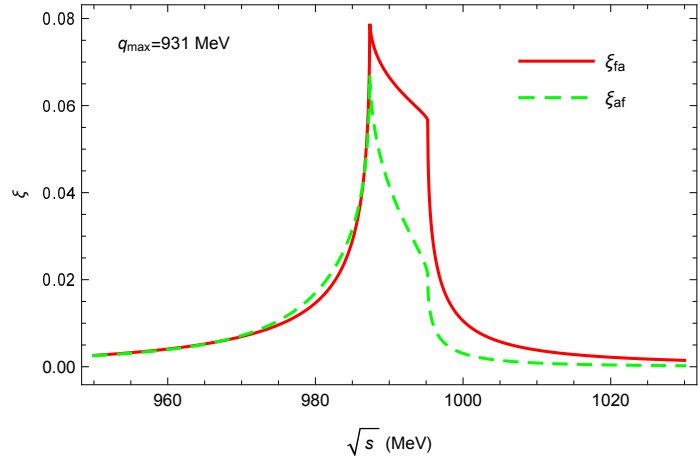}
\caption{The energy dependence of the mixing intensities $\xi_{fa}(s)$ (red solid line) and $\xi_{af}(s)$ (green dashed line). 
}
\label{fig:tmixint}
\end{figure}

Following Ref.~\cite{Ablikim:2018pik} we define
\begin{align}
\label{190601.1}
\xi_{fa}&=\frac{Br(J/\psi\to \phi f_0(980)\to \phi a_0(980)\to \phi\pi^0\eta)}
{Br(J/\psi\to \phi f_0(980)\to \phi \pi\pi)}~,\\
\xi_{af}&=\frac{Br(\chi_{c1}\to\pi^0a_0^0(980)\to\pi^0f_0(980)\to\pi^0\pi^+\pi^-)}{Br(\chi_{c1}\to\pi^0a_0^0(980)\to\pi^0\pi^0\eta)}~.
\end{align}
as an adequate way to measure the $f_0-a_0$ and $a_0-f_0$ mixing strengths, respectively.  

In connection with the previous definitions,
we also consider the energy-dependent mixing strengths $\xi_{fa}(s)$ and $\xi_{af}(s)$ defined as
\begin{align}
\label{190601.4}
\xi_{fa}(s)&=
\frac{|T_{35}(s)+T_{45}|^2 p_{\pi\eta}(s)}{3 |T_{32}(s)+T_{42}(s)|^2 p_{\pi\pi}(s)}~, \\
\xi_{af}(s)&=\frac{3 |T_{52}(s)|^2p_{\pi\pi}(s)}{|T_{55}(s)|^2p_{\pi\eta}(s)}~.\nn
\end{align}
The idea of the linear combinations of amplitudes in the first line of the previous equation
is to select the pure $I=0$ $K\bar{K}$ state in the sums $T_{35}+T_{45}$ and $T_{32}+T_{42}$, so that
the final decay into $\pi^0\eta$ is necessarily an isospin-breaking effect.
The dependence of these magnitudes with the energy around the $K\bar{K}$ thresholds is shown in
Fig.~\ref{fig:tmixint}, with $\xi_{fa}$ plotted by the (red) solid line and $\xi_{af}$ by the (green) dashed one. 

Now, we evaluate the ratios in Eq.~\eqref{190601.1} taking into account three-body phase space, see e.g. Ref.~\cite{pdg2018}.
We basically assume a constant coupling for the first vertex involving the heavy-meson particle and the scalar resonance.
In this way,
\begin{align}
\label{190601.5}
\xi_{fa}&=\frac{1}{3}\frac{\int dm_{12}^2 \int  dm_{23}^2 |T_{35}(m_{12}^2)+T_{45}(m_{12}^2)|^2}{\int dm_{12}^2 \int  dm_{23}^2 |T_{32}(m_{12}^2)+T_{42}(m_{12}^2)|^2}\\
=&\frac{1}{3}\frac{\int_{(m_1+m_2)^2}^{(M-m_3)^2}dm_{12}^2 \sqrt{E_2^{*2}-m_2^2}\sqrt{E_3^{*2}-m_3^2} |T_{35}(m_{12}^2)+T_{45}(m_{12}^2)|^2}{\int_{4m_\pi^2}^{(M-m_\pi)^2}dm_{12}^2 \sqrt{E_2^{*2}-m_\pi^2}\sqrt{E_3^{*2}-m_\pi^2} |T_{32}(m_{12}^2)+T_{42}(m_{12}^2)|^2}~,\nn
\end{align}
where $M=M_{J/\psi}$ and
\begin{align}
\label{190601.6}
E_2^{*}&=\frac{m_{12}^2-m_1^2+m_2^2}{2m_{12}}~,\\
E_3^{*}&=\frac{M^2-m_{12}^2-m_3^2}{2m_{12}}~.\nn
\end{align}
For the expression in the numerator $m_1=m_\pi$, $m_2=m_\eta$ and for the one in the denominator
$m_{1,2}=m_\pi$, with $m_3=m_\phi$ in both cases.

For the calculation of $\xi_{af}$ one has to take into account  the indistinguishability of the two
$\pi^0$ in the denominator of its definition.
In this case, $M=M_{\chi_{c1}}$, and for the decay in the
denominator $m_1=m_\pi$, $m_2=m_\eta$ and $m_3=m_\pi$, while for the one in the numerator $m_i=m_\pi$ for $i=1,2,3$.
With this preamble we have:
\begin{align}
\label{190601.7}
\xi_{af}&=\frac{\int dm_{12}^2\int dm_{23}^2 |T_{51}(m_{12}^2)|^2}{\frac{1}{2}\int dm_{12}^2\int dm_{23}^2 |T_{55}(m_{12}^2)+T_{55}(m_{23}^2)|^2}\nn\\
&=8\frac{\int_{4m_\pi^2}^{(M-m_\pi)^2}dm_{12}^2\sqrt{\tilde{E}_2^{*2}-m_\pi^2}\sqrt{\tilde{E}_3^{*2}-m_\pi^2}|T_{51}|^2}
{\int_{(m_\pi+m_\eta)^2}^{(M-m_\pi)^2} dm_{12}^2\int_{(m_{23}^2)_{{\rm min}}}^{(m_{23}^2)_{{\rm max}}} dm_{23}^2  |T_{55}(m_{12}^2)+T_{55}(m_{23}^2)|^2}~.
\end{align}
In the previous equation $\tilde{E}_2^*$ and $\tilde{E}_2^*$ correspond to $E_2^*$ and $E_3^*$ in Eq.~\eqref{190601.6}
with all $m_i=m_\pi$. On the other hand, for the denominator \cite{pdg2018}
\begin{align}
\label{190602.1}
(m_{23}^2)_{{\rm max}}&=(E_2^{*}+E_3^*)^2-(\sqrt{E_2^{*2}-m_2^2}-\sqrt{E_3^{*2}-m_3^2})^2~,\\
(m_{23}^2)_{{\rm min}}&=(E_2^{*}+E_3^*)^2-(\sqrt{E_2^{*2}-m_2^2}+\sqrt{E_3^{*2}-m_3^2})^2~,\nn
\end{align}
where $E_2^*$ and $E_3^*$ are given by Eq.~\eqref{190601.6} with the stated values for the different masses for this case.

Since one aims to isolate the signal of the $f_0(980)$ and $a_0(980)$ resonances  in the definitions of $\xi_{fa}$ and $\xi_{af}$,
respectively, we cut the PWAs in Eqs.~\eqref{190601.5} and \eqref{190601.7} such that $T_{ij}(s)\to T_{ij}(s)\theta(s_0-s)$,
with $s_0\simeq (1.2~\text{GeV})^2$.
This value of $s_0$ allows to fully cover the resonance region, while the PWAs from ChUA are still trustable for $s\lesssim s_0$.
The uncertainty given to the results because variations in the cutoff takes well into account  
reasonable changes in $s_0$ between $(1.1-1.2\,\text{GeV})^2$. We then obtain the values 
\begin{align}
\label{190703.1}
\xi_{fa}&=(8.9\pm 1.8)\times 10^{-3}~,\\
\xi_{af}&=(3.0\pm0.4)\times 10^{-3}~.\nn
\end{align}
It should be noted that these results are very compatible with the experimental ones reported
in Table~II of Ref.~\cite{Ablikim:2018pik}. 
Therefore, we dynamically generate the $f_0(980)-a_0(980)$ mixing effects in the coupled channel scattering amplitudes
and obtain consistent results with experiment for the ``mixing intensity''.
This outcome clearly favors the conclusion that the  $f_0(980)$ and $a_0(980)$ resonances have a strong dynamically generated component.

\section{The decays of the {\boldmath$J/\psi\to \gamma\eta\pi^0$, $J/\psi\to \gamma\pi^+\pi^-$} and  {\boldmath$J/\psi\to \gamma\pi^0\pi^0$}}

\begin{figure}
\centering
\includegraphics[scale=0.55]{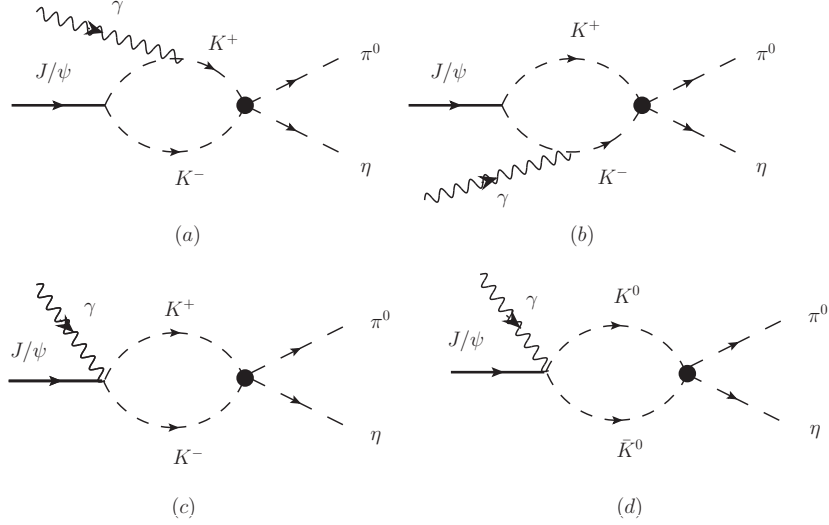}
\caption{Loop diagrams for the decay of $J/\psi\to \gamma\eta\pi^0$, where the black filled circle represents the full scattering amplitude
  from the kaons to the final state. }
\label{fig:diag2}
\end{figure}

Now we proceed to consider the resonance contributions of the $f_0(980)$ and $a_0(980)$ to the decays $J/\psi\to \gamma\eta\pi^0$,
$J/\psi\to \gamma\pi^+\pi^-$ and  $J/\psi\to \gamma\pi^0\pi^0$.
Inspired by Refs.~\cite{Oller:1998ia,Oller:2002na,Marco:1999df,Palomar:2003rb}, we employ the ChUA to study these decays.
Due to the  much higher thresholds of $D^+ D^-$ \cite{pdg2018} we do not include those channels in the triangle loops.
Besides, from the chiral Lagrangians considered, as discussed in Ref.~\cite{Marco:1999df,Bramon:1992ki}, there is no direct tree-level contribution
to the decays of the type $V \to P\, P\, \gamma$ (where $V$ and $P$ denote vector and pseudoscalar mesons, respectively). For a general discussion
of $VP$ Lagrangians, see e.g.~\cite{Meissner:1987ge}.
 Therefore, we take into account contributions
mediated by a $K^+K^-$ triangle loop with a photon line attached, as shown in the diagrams of Fig.~\ref{fig:diag2}. This mechanism is
also used for the other decays of $J/\psi\to \gamma\pi^+\pi^-$ and  $J/\psi\to \gamma\pi^0\pi^0$. 
The possibility of a contact-like contribution  to these decays involving the $K^0\bar{K}^0$ intermediate state, 
as represented in the last diagram of Fig.~\ref{fig:diag2}, is considered below. 
Notice that the final-state interactions are included by the resummation of infinite loops inherent to the ChUA, see Fig.~\ref{fig:diag}.

The amplitude for the radiative decay of $J/\psi\to \gamma\eta\pi^0$ in Fig. \ref{fig:diag2} can be written as
\begin{equation}
i{\cal M} = i\epsilon_{J/\psi}^\mu(P) \epsilon_\gamma^\nu(K) {\cal T}_{\mu\nu},
\end{equation}
where $P,\ K$ are the four momenta of the $J/\psi$ and $\gamma$, respectively. Since there are only two independent four-momenta,
the Lorentz covariant tensor ${\cal T}_{\mu\nu}$ can be written as
\begin{equation}
{\cal T}_{\mu\nu} = ag_{\mu\nu}+bP_\mu P_\nu+cP_\mu K_\nu+dP_\nu K_\mu+eK_\mu K_\nu.
\end{equation}
Taking into account that $P_\mu \epsilon_{J/\psi}^\mu=0$
and $\epsilon_\gamma^\nu K_\nu=0$,
only the two structures $a g_{\mu \nu}$ and $dP_\mu K_\mu$ survive.
In addition, because of gauge invariance, ${\cal T}_{\mu\nu}K^\nu=0$, and
therefore $a=-d K\cdot P$. In this way we can determine the function $a$ in terms of $d$, which is given by a convergent integral
\cite{Nussinov:1989gs}.
The $d$ coefficient from the triangle loops of the diagrams (a) and (b) in Fig.~\ref{fig:diag2} can be
evaluated using a Feynman parameterization of the corresponding loop function \cite{Nussinov:1989gs,Xiao:2012iq}, 
its analytical expression is given in Refs.~\cite{LucioMartinez:1990uw,Close:1992ay}, which we also
reproduce below.\footnote{Let us remark
  that a photon coupled to the right most vertex in Fig.~\ref{fig:diag2} does not give a contribution to the structure $K_\mu P_\nu$ because
  the loop integral only involves the total momentum $P$ in the denominator of the integrand. See Ref.~\cite{Oller:1998ia} for a more
  detailed discussion.}
For the diagram (c) in Fig.~\ref{fig:diag2}, a unitarity meson-meson loop appears, like those resummed by
the ChUA, although the associated subtraction constant does not need to be the same \cite{Oller:2002na}.
 The tree level amplitude for the decays $V \to P\, P\, \gamma$ is written analogously as the one for the same $\phi$ decays
as \cite{Close:1992ay} 
\begin{equation}
{\cal H}_{int} = (eA_\mu + g_{J/\psi} J/\psi_\mu)\; j^\mu - 2 e g_{J/\psi} A^\mu J/\psi_\mu K^{+\, \dagger} K^-, \label{eq:hint}
\end{equation}
where $A_\mu$, $J/\psi_\mu$ and $K^+ \, (K^-)$ are the photon, $J/\psi$ and charged kaon fields, $j^\mu
=i K^{+} (\overrightarrow{\partial^\mu}-\overleftarrow{\partial^\mu}) K^-$,
and the coupling $g_{J/\psi}$ can be evaluated from the decay width
\begin{equation}
\label{190512.1}  
\Gamma(J/\psi\to K^+ K^-) = \frac{g_{J/\psi}^2}{96 \pi} M_{J/\psi} \Big( 1-\frac{4m_{K^+}^2}{M_{J/\psi}^2} \Big)^2,
\end{equation}
where we add a factor of $1/\sqrt{2}$ for the field of $J/\psi_\mu$ which is analogous to the one of $\phi_\mu$ \cite{Klingl:1996by}.
Finally, we obtain the amplitude for Fig.~\ref{fig:diag2} as
\begin{equation}
{\cal M}_{\pi\eta} = -\sqrt{2}e \epsilon(J/\psi)\cdot\epsilon(\gamma) [g_{J/\psi} \widetilde{G}(M_{inv}^2) T_{K^+K^-\to \pi^0\eta}
+ g_c^{(I=1)} G(M_{inv}^2) T_{K\bar{K}\to \pi\eta}^{I=1}]~.
\label{eq:sumM}
\end{equation}
with $M_{inv}^2$  the invariant mass squared of the $\pi^0\eta$ state, $M_{inv}^2=(p_{\pi^0}+p_\eta)^2$. 
In the previous equation,  $g_c^{(I)}$ is an isospin dependent coupling ($I=0,1$) for the contact vertex  
  $V\gamma K\bar{K}$ already introduced in Ref.~\cite{Oller:2002na}.
The latter vertex is gauge invariant by itself, and we refer to this reference  for further details. 
This contributions stems from the diagrams (c) and (d) of Fig.~\ref{fig:diag2},
so that at the end the pure $I=1$ $|K\bar{K}\rangle$ state results
and couples strongly to the final state $|\pi\eta\rangle$ (which is purely $I=1$). 
 We also have the  functions $G(M_{inv}^2)$ and $\widetilde{G}(M_{inv}^2)$ in Eq.~\eqref{eq:sumM}.
The former is given by Eq.~\eqref{eq:Gco},\footnote{The Ref.~\cite{Oller:2002na} concludes that the unitarity-loop function $G(M_{inv}^ 2)$
  used in Eq.~\eqref{eq:sumM} could actually differ by a constant of its counterpart in the evaluation of the partial-wave amplitudes,
  cf. Eq.~\eqref{eq:Gco}. However, here we use a cutoff regularization for this function and insist on having a natural value for the cutoff
  around $1\gev$ in all the cases.}
and the latter results by the evaluation of the $K^+K^-\gamma$ triangle-loop graphs,
  and is given by \cite{LucioMartinez:1990uw,Close:1992ay}
\begin{eqnarray}
\widetilde{G}(M_{inv}^2) &=& \frac{1}{8\pi^2} (a-b) I(a,b), \ a=\frac{M_{J/\psi}^2}{M_K^2}, \ b=\frac{M_{inv}^2}{M_K^2},\\
  I(a,b) &=& \frac{1}{2(a-b)} - \frac{2}{(a-b)^2}\Big[ f\big(\frac{1}{b}\big) - f\big(\frac{1}{a}\big) \Big]
  + \frac{a}{(a-b)^2} \Big[ g\big(\frac{1}{b}\big) - g\big(\frac{1}{a}\big) \Big],\nonumber
\end{eqnarray}
with
\begin{eqnarray}
f(x)&=& \begin{cases}
-\Big[ \arcsin\big( \frac{1}{2\sqrt{x}} \big) \Big]^2, & x>\frac{1}{4}, \\
\frac{1}{4} \Big[ \log \big(\frac{\eta_+}{\eta_-}\big) - i \pi \Big]^2, & x<\frac{1}{4},  
\end{cases}  \\
g(x)&=& \begin{cases}
\Big[ \arcsin\big( \frac{1}{2\sqrt{x}} \big) \Big]^2 \sqrt{4x-1}, & x>\frac{1}{4}, \\
\frac{1}{2} \Big[ \log \big(\frac{\eta_+}{\eta_-}\big) - i \pi \Big]^2 \sqrt{1-4x}, & x<\frac{1}{4},  
\end{cases}  \\
\eta_{\pm} &=& \frac{1}{2x} \big[ 1\pm\sqrt{1-4x} \big].
\end{eqnarray}

Taking the modulus squared of Eq.~\eqref{eq:sumM}, summing over the polarizations of the photon
and averaging over those of the $J/\psi$, we get
\begin{equation}
\overline{\sum} \sum |{\cal M}_{\pi\eta}|^2 = \frac{4}{3} e^2 |g_{J/\psi} \widetilde{G}(M_{inv}^2) T_{K^+K^-\to \pi^0\eta} + g_c^{(I=1)}
G(M_{inv}^2) T_{K\bar{K}\to \pi\eta}^{I=1}]|^2,
\label{eq:sumM1}
\end{equation}
where the bar over the sum sign refers to the averaging process and in our normalization $e^2/4\pi \simeq 1/137$ is the fine structure constant.

By proceeding analogously as above to obtain Eq.~\eqref{eq:sumM}, we can write for the radiative $J/\psi$ decay to any of the two $\pi\pi$ modes
the following decay amplitude
\begin{equation}
\label{190510.3}
{\cal M}_{\pi_P\pi_Q} = -\sqrt{2}e \epsilon(J/\psi)\cdot\epsilon(\gamma) [g_{J/\psi} \widetilde{G}(M_{inv}^2) T_{K^+K^-\to \pi_P\pi_Q}
+ g_c^{(I=0)} G(M_{inv}^2) T_{K\bar{K}\to \pi_P\pi_Q}^{I=0} C_{PQ}], 
\end{equation}
where the Clebsch-Gordan coefficient $C_{PQ} = -\sqrt{2}/\sqrt{3}, \, -\sqrt{2}/\sqrt{6}$ for the $\pi^+\pi^-$,
and $\pi^0\pi^0$ case, respectively.\footnote{An extra factor of $\sqrt{2}$ is introduced in the Clebsch-Gordan coefficients
for the $\pi\pi$ states because of the unitarity normalization
for the $I=0$ $\pi\pi$ state~\cite{Oller:1997ti}.}
The sum over the final polarizations and the average over the initial ones of the
modulus squared of Eq.~\eqref{190510.3} gives an analogous expression to Eq.~\eqref{eq:sumM1}.

Besides, we use the isospin basis to evaluate the PWAs, as discussed before. We then have, 
\begin{equation}
T_{K^+K^-\to \pi^0\eta} = - \frac{1}{\sqrt{2}} T_{K\bar{K}\to \pi\eta}^{I=1},
\end{equation}
 whereas for the other two cases we have
\begin{eqnarray}
T_{K^+K^-\to \pi^+\pi^-} &= \frac{1}{\sqrt{3}} T_{K\bar{K}\to \pi\pi}^{I=0}, \\
T_{K^+K^-\to \pi^0\pi^0} &= \frac{1}{\sqrt{6}} T_{K\bar{K}\to \pi\pi}^{I=0}, \label{eq:tkpi}
\end{eqnarray}
where there is a factor of $1/\sqrt{2}$ to account for the identity of the two neutral pions.

With the amplitudes obtained  above, the differential decay widths of the $J/\psi$ can be calculated by 
\begin{equation}
\frac{d\Gamma_{\gamma PQ}}{dM_{inv}} = \frac{1}{(2 \pi)^3} \frac{1}{4 M_{J/\psi}^2} \, p_\gamma \, \tilde{p}_\eta \, \overline{\sum} \sum |{\cal M}|^2,
\label{190510.4}
\end{equation}
where
\begin{eqnarray}
p_\gamma &= \frac{\lambda^{1/2}(M_{J/\psi}^2, M_{inv}^2, 0)}{2 M_{J/\psi}} , \\
\tilde{p}_\eta &= \frac{\lambda^{1/2}(M_{inv}^2, M_\eta^2, M_{\pi^0}^2)}{2M_{inv}}, 
\end{eqnarray} 
 $\lambda(a,b,c)=a^2+b^2+c^2-2(ab+ac+bc)$ is the usual K\"allen triangle function, and $p_\gamma$, $\tilde{p}_\eta$ are the $\gamma$ momentum
in the ${J/\psi}$ rest-frame and the $\eta$ momentum in the $\pi^0\eta$ rest-frame, respectively.
The integration of Eq.~\eqref{190510.4} allows us to  calculate the decay widths, 
\begin{equation}
\Gamma_{\gamma PQ} = \int \frac{d\Gamma_{\gamma PQ}}{dM_{inv}} \,dM_{inv}.
\label{eq:gam}
\end{equation}

The numerical values taken for the masses of the particles are  \cite{pdg2018}:
$M_{K^+}=M_{K^-}=493.677\mev$, $M_{K^0}=497.611\mev$, $M_{\pi^+}=M_{\pi^-}=139.57018\mev$,
$M_{\pi^0}=134.9766\mev$, $M_{\eta}=547.862\mev$ and $M_{J/\psi}=3096.9\mev$.
With the branching ratio ${\rm Br}(J/\psi\to K^+K^)=2.86 \times 10^{-4}$ 
and the total width $\Gamma_{J/\psi}=92.9 \times 10^{-3} \mev$ \cite{pdg2018},
we get from Eq.~\eqref{190512.1} the coupling $g_{J/\psi} = 1.74 \times 10^{-3}$.
Since the $J/\psi$ is a $c\bar{c}$,
one can expect that the $K\bar{K}$ stems from a $0^{++}$ source in the direct coupling $J/\psi\gamma K\bar{K}$. 
The reason is the same as advocated in Ref.~\cite{Meissner:2000bc}
for the study of the $J/\psi \to \phi \pi\pi$ decay, 
where in our case the $\phi$ (a vector) is playing the analogous role to the photon.
Therefore, one would expect that the $K\bar{K}$ is in $I=0$,  and then we take in the following that $g_c^{(I=1)}=0$,
though we keep it in the algebraic equations.\footnote{The $J/\psi$ is a $c\bar{c}$ and its decay into 
light-quark hadrons is an isoscalar OZI violating process rich in intermediate gluons that we also take as a scalar 
source following Ref.~\cite{Meissner:2000bc}.
This is similar to the well-known $^3P_0$ decay model in the quark model \cite{Micu:1968mk} where a $q\bar{q}$
is produced from the vacuum and having its same quantum numbers.
In this way, the isoscalar part of the electromagnetic current is selected
because the radiative coupling of the photon to the charge kaons is already accounted for
by the diagrams (a) and (b) of Fig.~\ref{fig:diag2}.}
By performing the isospin decomposition we then have that $g_{J/\psi \gamma K^+K^-}=-g_c^{(I=0)}/\sqrt{2}$. 
For some estimations below we make the identification  $g_c^{(I=0)} \to -\sqrt{2} g_{J/\psi}$,
and discuss its uncertainties afterwards.

\begin{figure}
\centering
\includegraphics[scale=0.8]{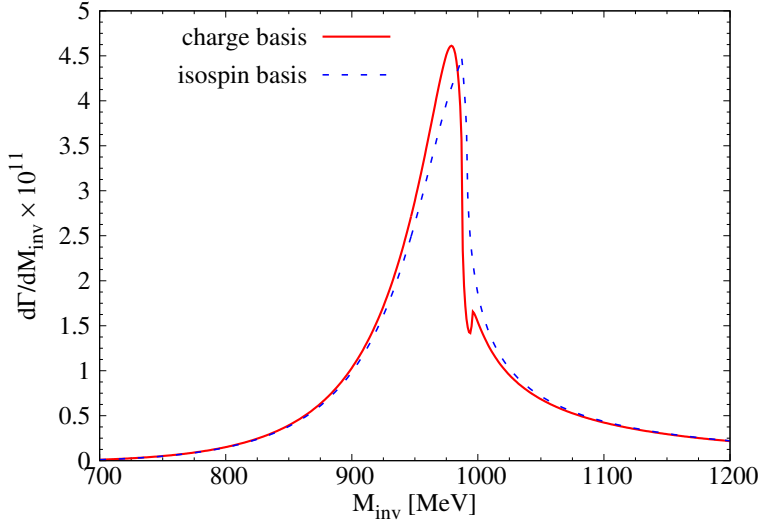}
\caption{The $\eta\pi^0$ invariant mass distribution for the decay width of $J/\psi\to \gamma\eta\pi^0$.
}
\label{fig:res1}
\end{figure}
We show the $\eta\pi^0$ invariant mass distribution for the decay $J/\psi \to \gamma \pi^0\eta$  
in Fig.~\ref{fig:res1} from threshold up to around 1.2~GeV, so that the $a_0(980)$  resonance signal is fully covered.
Let us indicate that now the final state is purely $I=1$ and then $g_c^{(I=1)}$ is the one entering in the calculations which is taken to be zero, 
as discussed above. 
Therefore, the transition amplitude in Eq.~\eqref{eq:sumM} is fixed in this case within our model calculation.
The results obtained by employing the isospin or charge bases of states for the strong interactions are shown by the 
  dashed and solid lines, respectively. They are consistent with each other, 
as expected from our considerations above about the calculation of the strong amplitudes in one basis or the other.
By integrating the invariant mass distribution calculated we obtain
$Br(J/\psi\to \gamma a_0(980) \to  \gamma\eta\pi^0) = 0.48 \times 10^{-7}$,
  with the resonance contributions of the $a_0(980)$ accounted for.
The BESIII Collaboration reports in \cite{Ablikim:2016exh} the branching fraction
$Br(J/\psi\to \gamma\eta\pi^0) = 2.14 \times 10^{-5}$ and an upper limit for the $a_0(980)$
intermediate contribution of $Br(J/\psi\to \gamma a_0(980) \to  \gamma\eta\pi^0) = 2.5 \times 10^{-6}$.
Thus, our results is an order of magnitude smaller than this upper bound and consistent with the BESIII measurements.
Besides, our estimate for the $a_0(980)$ contributions are two orders of magnitude smaller than the branching
fraction of $J/\psi\to \gamma\eta\pi^0$, which indicates that other mechanisms apart from the exchange of the $a_0(980)$
  dominate this decay process, like e.g. higher resonances or contact terms,
with the $a_2(1320)$ explicitly found in Ref. \cite{Ablikim:2016exh} as well.

\begin{figure}
\centering
\includegraphics[scale=0.6]{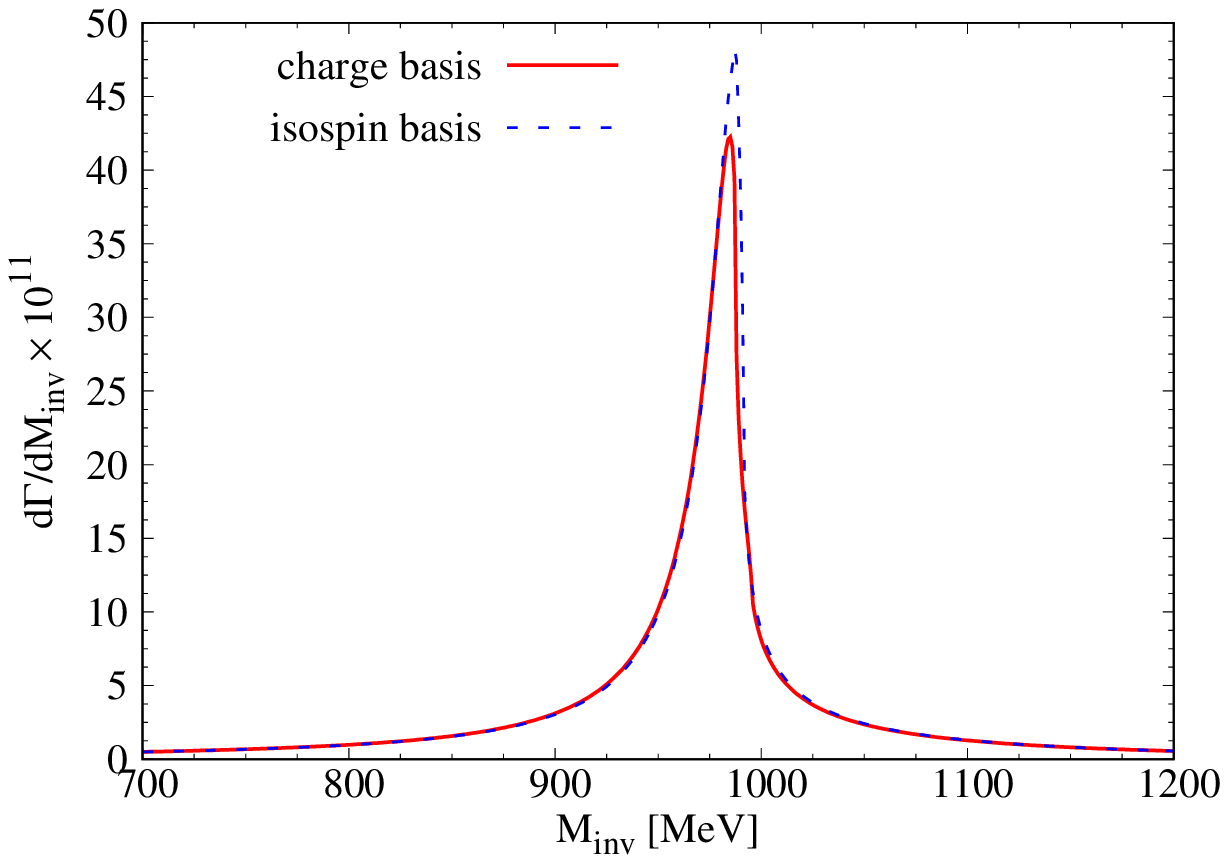}
\includegraphics[scale=0.6]{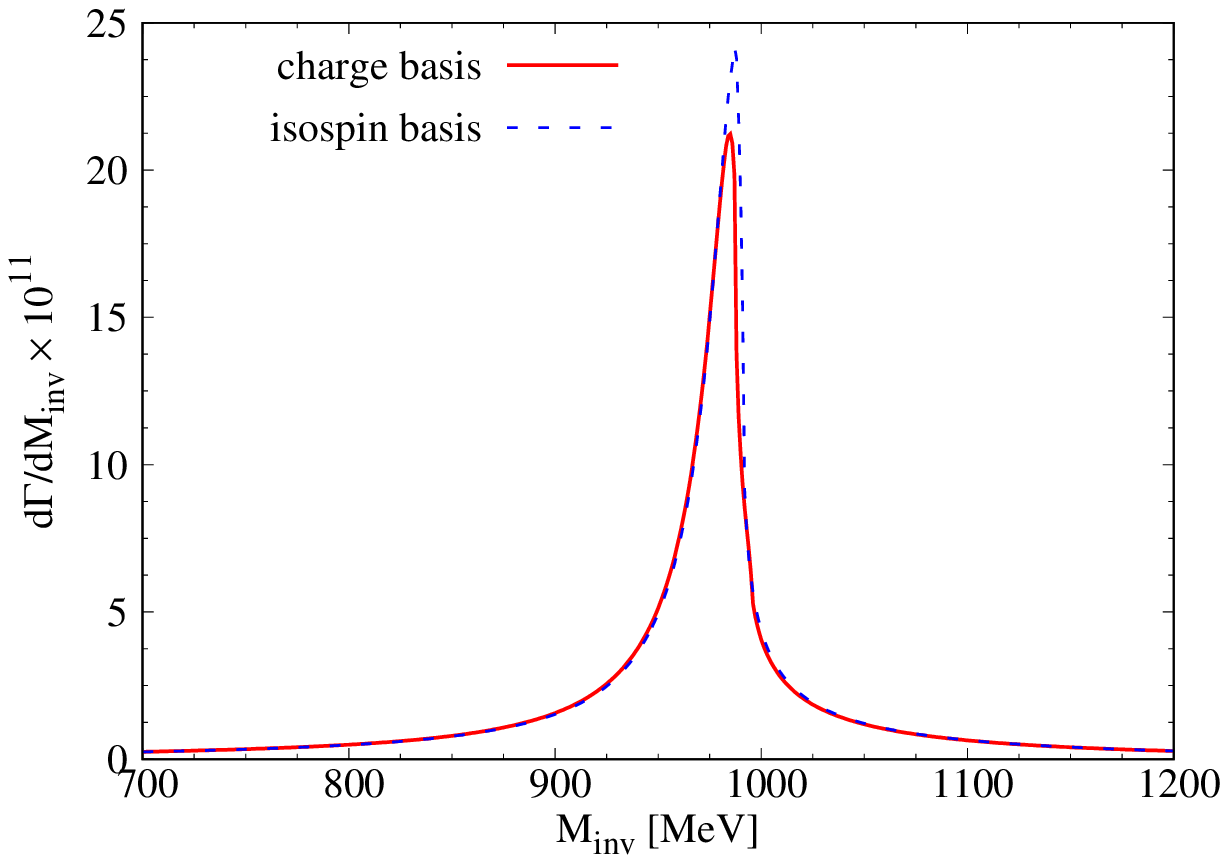}
\caption{The $\pi\pi$ invariant mass distributions for the partial decay widths 
$J/\psi\to \gamma\pi^+\pi^-$ (left panel) and  $J/\psi\to \gamma\pi^0\pi^0$ (right panel). 
To draw the figures $g_c^{(I=0)}=-\sqrt{2}g_{J/\psi}$ is taken for definiteness.
}
\label{fig:res2}
\end{figure}

We show in Fig.~\ref{fig:res2} the $\pi\pi$ invariant  mass distributions for
the decays $J/\psi\to \gamma\pi^+\pi^-$ and  $J/\psi\to \gamma\pi^0\pi^0$, which only differ by a factor of 2 
because of the the remark below Eq. \eqref{eq:tkpi}.
For illustrative purposes we have taken that $g_c^{I=0}=-\sqrt{2} g_{J/\psi}$ to obtain the curves in the figure.
The branching fractions obtained for these two processes are
$Br(J/\psi\to \gamma f_0(980) \to  \gamma\pi^+\pi^-) = 2.39 \times 10^{-7}$
and $Br(J/\psi\to \gamma f_0(980) \to  \gamma\pi^0\pi^0) = 1.20 \times 10^{-7}$.
The BESIII Collaboration has studied the decay of $J/\psi\to \gamma\pi^0\pi^0$
in Ref. \cite{Ablikim:2015umt} within an amplitude analysis
and determined the branching ratio $Br(J/\psi\to \gamma\pi^0\pi^0) = 1.15 \times 10^{-3}$.
From the Fig.~2(a) of Ref. \cite{Ablikim:2015umt}, we can see that the dominant contributions stem
from the higher resonances, such as the $f_0(1370)$, $f_0(1500)$, $f_0(1710)$,
and just a possible structure of the $f_0(980)$, with a quite small enhancement not observed in earlier
measurements \cite{Ablikim:2006db}.
Since the scanned energy  in Ref. \cite{Ablikim:2006db} is only above $1\gev$, only higher resonance contributions
are found in these decays.
The $f_0(980)$ state is also notably absent from the $J/\psi$ radiative decay
in the early experimental observations \cite{Becker:1986zt,Augustin:1987da,Bai:1996wm}.
These observations are in agreement with our results which lead to very 
small contributions of the  $a_0(980)$ and $f_0(980)$ resonances to the branching fractions 
$J/\psi\to \gamma\eta\pi^0$, $J/\psi\to \gamma\pi^+\pi^-$ and  $J/\psi\to \gamma\pi^0\pi^0$. 
The large mass of the $J/\psi$ particle also favors this situation because it
leads to plenty of available phase space for the higher resonances to contribute, as found in Ref. \cite{Ablikim:2018izx}.
Furthermore, comparing the results of Fig. \ref{fig:tsqf0a03} with the ones in Fig. \ref{fig:tsqf0a02},
the $f_0(980)-a_0(980)$ mixing effects are very difficult to observe in these $J/\psi$ radiative decays
because of the relative weakness of the mixing amplitudes $|T_{15}|^2$ and $|T_{25}|^2$.
This is within expectations because the external photon probe already involves $I=0,\,1$, so that the isospin-conserving
contributions in the final-state interactions dominate.

\begin{figure}
\centering
\includegraphics[scale=0.6]{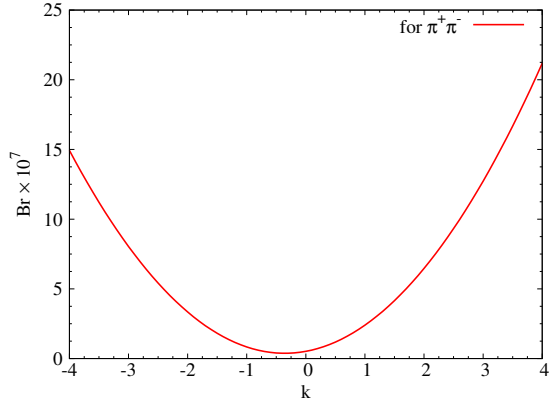}
\caption{The branching ratio for $J/\psi\to \gamma\pi^+\pi^-$ as a function of
$k=-g_c^{(I=0)}/\sqrt{2}g_{J/\psi}$.}
\label{fig:bradep}
\end{figure}

In order to see the dependence of our results with the unknown parameter $g_c^{(I=0)}$ we plot in Fig.~\ref{fig:bradep} how 
  the calculated $Br(J/\psi\to \gamma f_0(908)\to \gamma \pi^+\pi^-)$ 
  depends on the ratio $k=g_c^{(I=0)}/(-\sqrt{2} g_{J/\psi})$. For large enough $k$ we observe a parabolic dependence because
the decay width scales then as $g_c^{(I=0)}$ squared, cf. Eq.~\eqref{190510.3}.
The most stable results under changes of the coupling $g_c^{(I=0)}$ occurs in the range for $k\in[-1,0]$, with a range of values 
 $Br(J/\psi\to\gamma \pi^+\pi^-)\sim (0.5-1)\times 10^{-7}$, and a half of it for the neutral mode $J/\psi\to \gamma\pi^0\pi^0$.

Regarding the dependence of our results on  the cutoff $q_{\rm max}$,
we estimate the uncertainty associated to changing this parameter within a 10\%, which is the
range of values that we concluded when fitting the experimental data in Sec.~\ref{sec.190501.1}.

Taking into account the theoretical uncertainties in the cutoff, we obtain a rather definite 
prediction for the $Br(J/\psi\to \gamma a_0(980) \to \gamma\pi^0\eta)$, 
\begin{align}
\label{190513.1}
Br(J/\psi\to \gamma a_0(980) \to \gamma\eta\pi^0) = (0.48\pm 0.03) \times 10^{-7}~. 
\end{align}

The situation is of course much more uncertain for the  $J/\psi\to \gamma \pi\pi$ branching ratios
because of the uncertainty in the unknown parameter $g_c^{(I=0)}$.
Nonetheless, there is a priori no reason to conclude that the contribution proportional to $g_c^{(I=0)}$ should be much larger
than the one proportional to $g_{J/\psi}$ from our calculation of the $M_{\pi_P\pi_Q}$ decay amplitudes in Eq.~\eqref{190510.3}.
Indeed, one would expect precisely the opposite situation because the relation of
$\widetilde{G}(s)$ with the derivative of the unitarity function $G(s)$ with respect to the energy at around the
two-kaon threshold, where the latter function has a branch-point singularity.
As a result, we take as interval of values for our estimation for the $\gamma\pi\pi$ branching ratios the one comprised by
the absolute minimum shown in Fig.~\ref{fig:bradep} and four times its value for $k=0$.
This implies to take the quite conservative attitude of allowing  the 
decay amplitude ${\cal M}_{\pi_Q\pi_Q}$ to double its size with respect to $k=0$ because of having included
the term proportional to $g_c^{(I=0)}$ in Eq.~\eqref{190510.3}. We then take the interval of values,
\begin{align}
\label{190513.2}
Br(J/\psi\to \gamma f_0(980) \to  \gamma\pi^+\pi^-) &= (0.52-2.08) \times 10^{-7}~, \\
Br(J/\psi\to \gamma f_0(980) \to  \gamma\pi^0\pi^0) &= (0.26-1.04) \times 10^{-7}~. \nonumber
\end{align}

\begin{figure}
\centering
\includegraphics[scale=0.55]{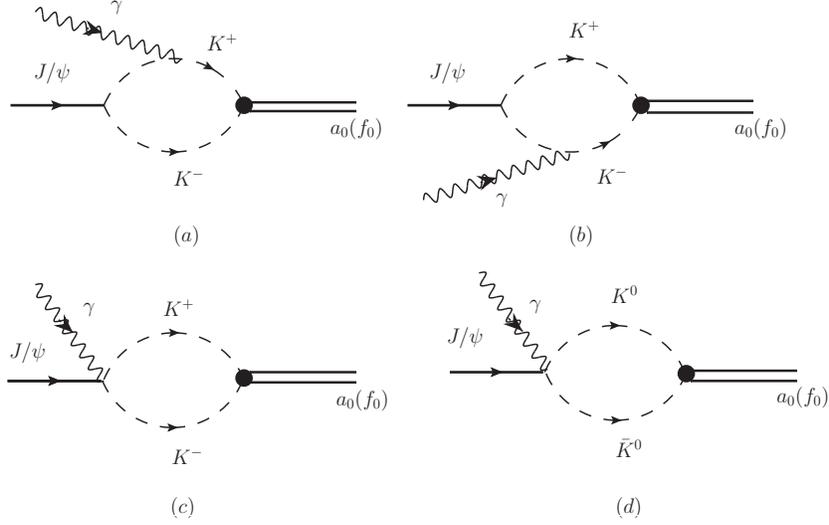}
\caption{Loop diagrams for the decay  $J/\psi\to \gamma a_0(f_0)$ by isolating the resonance pole in the diagrams 
of Fig.~\ref{fig:diag2}.}
\label{fig:diag3}
\end{figure}

We can also obtain from our results the related branching ratios $Br(J/\psi\to\gamma f_0(980))$ and
  $Br(J/\psi\to\gamma f_0(980))$ 
by  considering directly the production of the resonances $a_0(980)$ and $f_0(980)$
as final states in the radiative decays, see  Fig.~\ref{fig:diag3}.
Proceeding analogously as in Eqs.~\eqref{eq:sumM}, \eqref{eq:sumM1} and \eqref{190510.3}, we readily obtain
\begin{equation} 
\overline{\sum} \sum |{\cal M}_R|^2 = \frac{4}{3} e^2 |g_{J/\psi} \widetilde{G}(M_{inv}^2) g^{(I)}_{RK^+K^-}
+ g_c^{(I)} G(M_{inv}^2) (-\sqrt{2}g^{(I)}_{RK^+K^-})|^2,
\label{eq:sumM2}
\end{equation}
where $g^{(I)}_{RK^+K^-}$ is the coupling to $K^+K^-$ of the resonance $f_0(980)$ ($I=0$) or $a_0(980)$ ($I=1$).
Notice that $g^{(I)}_{RK\bar{K}} =- \sqrt{2} g^{(I)}_{RK^+K^-}$ for both resonances, where $g^{(I)}_{RK\bar{K}}$ is the
coupling to the $K\bar{K}$ with definite isospin.
We also identify the invariant mass with the resonance mass, so that $M_{inv} = m_{a_0(f_0)}$.
Thus, the radiative decay widths of the $J/\psi$ to these resonances are given by 
\begin{equation}
\Gamma_{\gamma a_0(f_0)} =\frac{p_\gamma}{8 \pi M_{J/\psi}^2} \,  \overline{\sum} \sum |{\cal M}_R|^2, \label{eq:gam2}
\end{equation}
with $p_\gamma$ the three momentum of the photon in the rest frame of the $J/\psi$ particle.
The value of the coupling $g^{(I)}_{RK^+K^-}$ can  be calculated straightforwardly from the ChUA amplitudes 
already obtained in Sec.~\ref{sec.190501.1} \cite{Oller:1998zr}.

Our results are sharper for the $Br(J/\psi\to \gamma a_0(980))$ by having argued that
$g_c^{(I=1)}=0$, cf. the discussion after  Eq.~\eqref{eq:gam}, with the result 
\begin{align}
\label{190513.3}
Br(J/\psi\to \gamma a_0(980)) = (1.24 \sim 1.61) \times 10^{-7}.
\end{align}
where the uncertainty arises from the variation in the cutoff.

In turn, the uncertainty for the $Br(J/\psi\to \gamma f_0(980))$ are much larger and we follow the same idea as explained
with respect to Eq.~\eqref{190513.2}, so as to make an estimation of the uncertainty associated to the unknown parameter $g_c^{I=0}$.
In this way, we calculate the minimum value of the branching ratio as a function of $g_c^{(I=0)}$ and multiply  
by four the resulting one for $g_c^{(I=0)}=0$. In this way, we obtain the interval of values 
\begin{align}
\label{190513.4}
Br(J/\psi\to \gamma f_0(980)) = (0.69 \sim 4.00) \times 10^{-7}~,
\end{align}
where in addition we have taken into account the uncertainty in $q_{\rm max}$.

\begin{figure}
\centering
\includegraphics[scale=0.6]{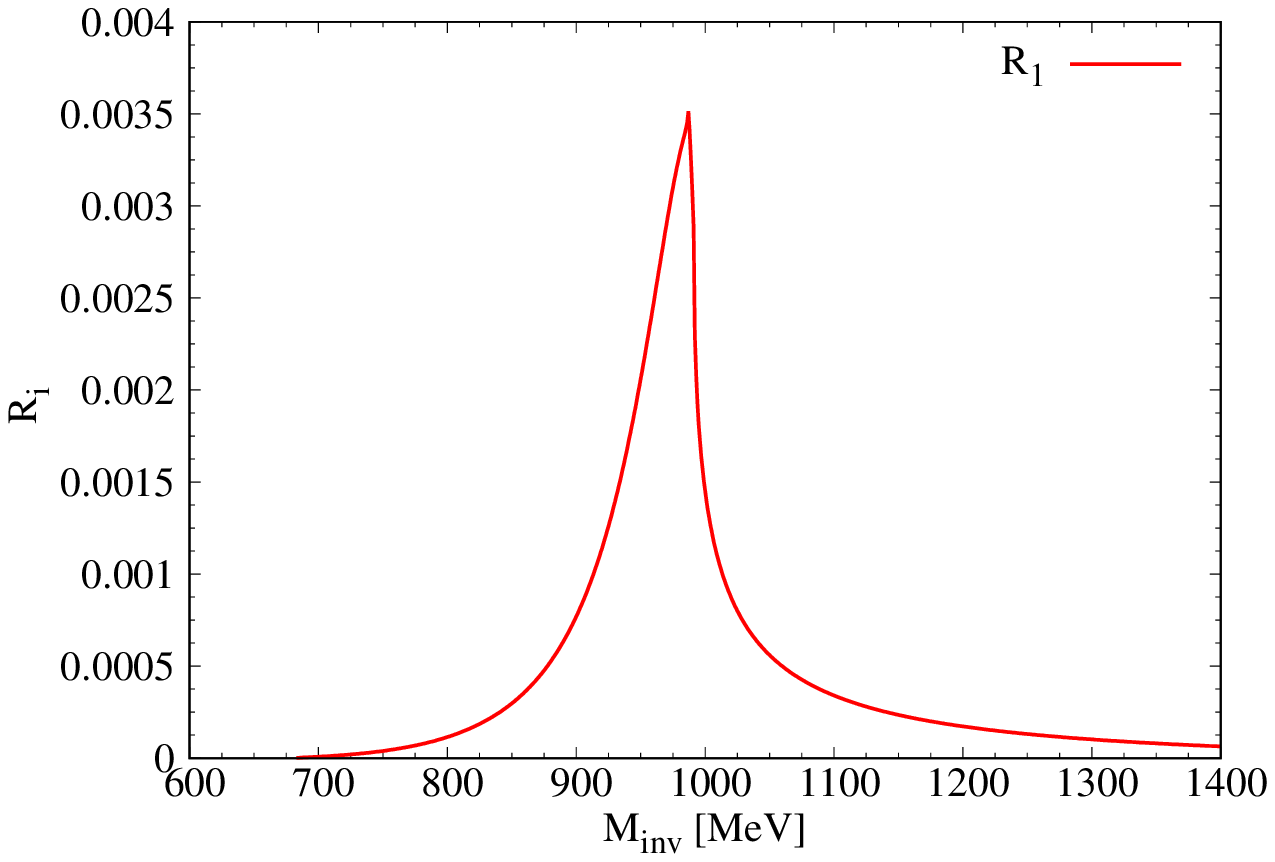}
\includegraphics[scale=0.6]{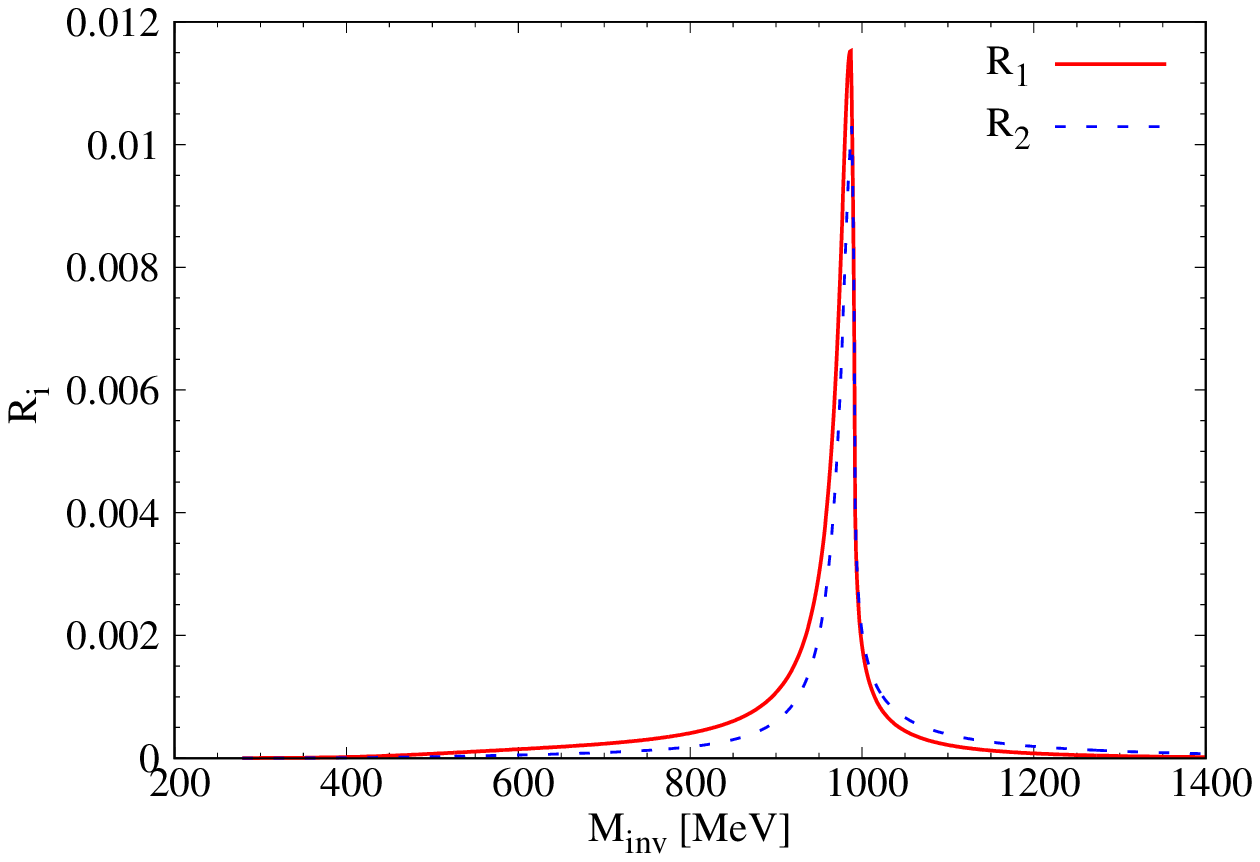}
\caption{The results of the ratios $R_1$ and $R_2$ for the $ a_0(980)$ (left) and  $f_0(980)$ (right) resonances, 
  where we have taken $g_{a_0 K^+K^-}=3.60\gev$ and  $g_{f_0 K^+K^-} = 2.78\gev$.}
\label{fig:res3}
\end{figure}

One further output from  our approach follows by defining the ratios,
\begin{equation}
\label{190513.5}  
R=\Gamma_{\gamma PQ}/\Gamma_{\gamma a_0(f_0)},
\end{equation}
as in Refs.~\cite{MartinezTorres:2012du,Xie:2015lta}. 
In this way, one can remove the dependence on $g_{J/\psi}$ (recall that $g_c^{(I=1)}=0$) in the ratio for the $a_0(980)$.
However, for the $f_0(980)$ this is not the case because $g_c^{(I=0)}\neq 0$ and both couplings add coherently then.
  In addition, we also introduce the ratios
  \begin{align}
    \label{190601.2}
 R_1&=\frac{d\Gamma_1 / dM_{inv}}{\Gamma_{\gamma a_0(f_0)\;1}}~,\\
 R_2&=\frac{d\Gamma_2 / dM_{inv}}{\Gamma_{\gamma a_0(f_0)\;2}}~, \\
 R_1'&=\frac{\Gamma_1}{\Gamma_{\gamma a_0(f_0)\,1}}~,\\
 R_2'&=\frac{\Gamma_2}{\Gamma_{\gamma a_0(f_0)\,2}}~, 
\end{align}
where $d\Gamma_1 / dM_{inv}$, $d\Gamma_2 / dM_{inv}$  and $\Gamma_{\gamma a_0(f_0)\,1}$, $\Gamma_{\gamma a_0(f_0)\,2}$
come from the two parts contributing to the decay amplitudes squared, cf. Eqs.~\eqref{eq:sumM1} and \eqref{eq:sumM2}.
Namely, 
\begin{align}
&\frac{d\Gamma_1}{dM_{inv}} \to \frac{4}{3} e^2 |g_{J/\psi} \widetilde{G}(M_{inv}^2) T_{K^+K^-\to \pi^0\eta(\pi\pi)}|^2, \\
&\frac{d\Gamma_2}{dM_{inv}} \to \frac{4}{3} e^2 |g_c^{(I)} G(M_{inv}^2) T_{K\bar{K}\to \pi\eta(\pi\pi)}^{I=1(0)} C_{PQ}|^2, \\
&\Gamma_{\gamma a_0(f_0)\,1} \to \frac{4}{3} e^2 |g_{J/\psi} \widetilde{G}(M_{inv}^2) g_{RK^+K^-}|^2, \\
&\Gamma_{\gamma a_0(f_0)\,2} \to \frac{8}{3} e^2 |g_c^{(I)} G(M_{inv}^2) g_{RK^+K^-}|^2.
\end{align}
For the $a_0(980)$ case, we obtain $R=R'_2=0.28\pm 0.41$.
 Taking $k=g_c^{(I=0)}/(-\sqrt{2} g_{J/\psi})=1$ we calculate for the $f_0(980)$ resonance that $R=0.46-0.54$, 
$R_1'=0.57-0.65$ and $R_2'=0.46-0.54$.
 The results for $R_1$ and $R_2$ are shown in Fig. \ref{fig:res3}
as a function of the invariant mass squared of the pair of pseudoscalars.
There is not $R_2$ for the $a_0(980)$ resonance case because $g_c^{(I=1)}=0$. 
In the plots we take $g_{a_0 K^+K^-} = 3.60\gev$ and $g_{f_0 K^+K^-} = 2.78\gev$ for convenience
since the results for different couplings are only affected by a factor.
In Fig. \ref{fig:res3}, the contributions of the $ a_0(980)$ and  $f_0(980)$ resonances can be clearly seen.

\begin{figure}
\centering
\includegraphics[scale=0.6]{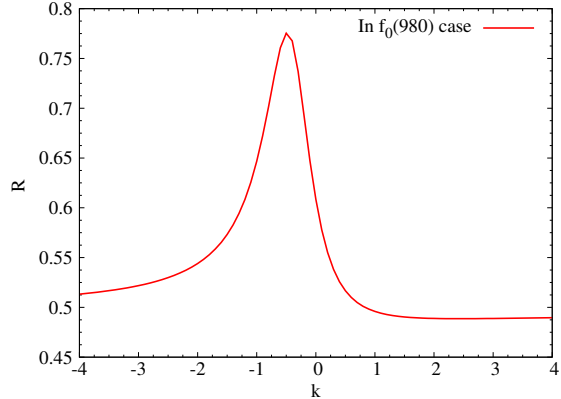}
\caption{The dependence of the ratio $R$ with the coupling $g_c^{(I=0)}$, 
where we have taken $g_{f_0 K^+K^-} = 2.78\gev$.}
\label{fig:ratdep}
\end{figure}

In addition,  we show in Fig. \ref{fig:ratdep} the dependence of the ratio $R$
on the unknown coupling constant $g_c^{(I=0)}$ with the $x$ axis corresponding to
$k=g_c^{(I=0)}/(-\sqrt{2} g_{J/\psi})$. In the figure we have taken $g_{f_0 K^+K^-} = 2.78\gev$.
Let us notice that  for $k>0.5$ one enters in a plateau region for $R$ in the case of the $f_0(980)$, 
with a value around 0.49.

\section{Conclusions}
\label{sec.190501.3}

In the present work, we have revisited the $K\bar{K}$ interactions in coupled channels,
where we have dynamically reproduced the resonances of $f_0(980)$ and $a_0(980)$
with both the isospin-basis and the charge-basis formalisms.
Interestingly, we can dynamically generate the $f_0(980)-a_0(980)$ mixing effects
in the scattering amplitudes of the coupled channels
with the charge basis, within the assumption of a dominant hadronic nature for these resonances.
One can also easily estimate the ``mixing intensities'' $\xi_{fa}$ and $\xi_{af}$
around the energy range of the resonances within this framework,
and values in remarkable agreement with the experimental results of Ref.~\cite{Ablikim:2018pik} are obtained.

With the scattering amplitudes of the coupled channels reproduced, we investigate
the contributions through the final state interactions
of the resonances $f_0(980)$ and $a_0(980)$ to the $J/\psi$ radiative decays 
$J/\psi\to \gamma\eta\pi^0$, $J/\psi\to \gamma\pi^+\pi^-$ and  $J/\psi\to \gamma\pi^0\pi^0$.
Taking into account the theoretical uncertainties associated to a 10\% of variation in the three-momentum
cutoff $q_{{\rm max}}$,
we obtain the branching fraction
\begin{align*}
&Br(J/\psi\to \gamma a_0(980) \to  \gamma\eta\pi^0) = (0.48\pm0.03) \times 10^{-7}, \\
\end{align*}
The uncertainties affecting the $Br(J/\psi\to \gamma\pi\pi)$ are larger because their dependence on the unknown
  coupling $g_c^{(I=0)}$. We then end with the interval of values
\begin{align*}
&Br(J/\psi\to \gamma f_0(980) \to  \gamma\pi^+\pi^-) = (0.52-2.08) \times 10^{-7}, \\
&Br(J/\psi\to \gamma f_0(980) \to  \gamma\pi^0\pi^0) = (0.26-1.04) \times 10^{-7}.
\end{align*}
In terms of the same formalism one can also predict
\begin{align*}
&Br(J/\psi\to \gamma a_0(980)) =  (1.24 \sim 1.61) \times 10^{-7},\\
&Br(J/\psi\to \gamma f_0(980)) = (0.69 \sim 4.00) \times 10^{-7}.
\end{align*}
These fractions are within the upper limits for these decays from the experimental measurements.
The small results that we have found for the contributions of the $a_0(980)$ and $f_0(980)$ resonances
in these decays agree with the dominant contributions from higher resonances as found experimentally.
From these results, we conclude that it is more difficult to detect the $f_0(980)-a_0(980)$ mixing effects
in these radiative decay channels of the $J/\psi$ particle.
We expect that future measurements within  higher statistical experiments can support our results.

\section*{Acknowledgments}

CWX thanks E. Oset and Q. Wang for the useful discussions.
This work is supported in part by the DFG and the NSFC through
funds provided to the Sino-German CRC~110 ``Symmetries and
the Emergence of Structure in QCD''.
JAO would like to thank partial financial support by the MINECO (Spain) and FEDER (EU) grant FPA2016-77313-P.
The work of UGM was also supported by by the Chinese  Academy of Sciences (CAS) President's International Fellowship Initiative (PIFI)
(grant no. 2018DM0034)  and by VolkswagenStiftung (grant no. 93562).


\begin{thebibliography}{99}



  
\bibitem{Aaij:2015tga} 
  R.~Aaij {\it et al.} [LHCb Collaboration],
  Phys.\ Rev.\ Lett.\  {\bf 115}, 072001 (2015)
  [arXiv:1507.03414 [hep-ex]].

\bibitem{Aaij:2019vzc} 
  R.~Aaij {\it et al.} [LHCb Collaboration],
  arXiv:1904.03947 [hep-ex].

\bibitem{Chen:2016qju} 
  H.~X.~Chen, W.~Chen, X.~Liu and S.~L.~Zhu,
  Phys.\ Rept.\  {\bf 639}, 1 (2016)
  [arXiv:1601.02092 [hep-ph]].

\bibitem{Hosaka:2016pey} 
  A.~Hosaka, T.~Iijima, K.~Miyabayashi, Y.~Sakai and S.~Yasui,
  PTEP {\bf 2016}, no. 6, 062C01 (2016)
  [arXiv:1603.09229 [hep-ph]].

\bibitem{Chen:2016spr} 
  H.~X.~Chen, W.~Chen, X.~Liu, Y.~R.~Liu and S.~L.~Zhu,
  Rept.\ Prog.\ Phys.\  {\bf 80}, no. 7, 076201 (2017)
  [arXiv:1609.08928 [hep-ph]].

\bibitem{Lebed:2016hpi} 
  R.~F.~Lebed, R.~E.~Mitchell and E.~S.~Swanson,
  Prog.\ Part.\ Nucl.\ Phys.\  {\bf 93}, 143 (2017)
  [arXiv:1610.04528 [hep-ph]].

\bibitem{Esposito:2016noz} 
  A.~Esposito, A.~Pilloni and A.~D.~Polosa,
  Phys.\ Rept.\  {\bf 668}, 1 (2016)
  [arXiv:1611.07920 [hep-ph]].

\bibitem{Guo:2017jvc} 
  F.~K.~Guo, C.~Hanhart, U.~G.~Mei{\ss}ner, Q.~Wang, Q.~Zhao and B.~S.~Zou,
  Rev.\ Mod.\ Phys.\  {\bf 90}, no. 1, 015004 (2018)
  [arXiv:1705.00141 [hep-ph]].

\bibitem{Ali:2017jda} 
  A.~Ali, J.~S.~Lange and S.~Stone,
  Prog.\ Part.\ Nucl.\ Phys.\  {\bf 97}, 123 (2017)
  [arXiv:1706.00610 [hep-ph]].

\bibitem{Olsen:2017bmm} 
  S.~L.~Olsen, T.~Skwarnicki and D.~Zieminska,
  Rev.\ Mod.\ Phys.\  {\bf 90}, no. 1, 015003 (2018)
  [arXiv:1708.04012 [hep-ph]].

\bibitem{Karliner:2017qhf} 
  M.~Karliner, J.~L.~Rosner and T.~Skwarnicki,
  Ann.\ Rev.\ Nucl.\ Part.\ Sci.\  {\bf 68}, 17 (2018)
  [arXiv:1711.10626 [hep-ph]].

\bibitem{Yuan:2018inv} 
  C.~Z.~Yuan,
  Int.\ J.\ Mod.\ Phys.\ A {\bf 33}, no. 21, 1830018 (2018)
  [arXiv:1808.01570 [hep-ex]].

\bibitem{Astier:1967zz} 
  A.~Astier, L.~Montanet, M.~Baubillier and J.~Duboc,
  Phys.\ Lett.\  B {\bf 25}, 294 (1967).

\bibitem{Ammar:1969vy} 
  R.~Ammar {\it et al.},
  Phys.\ Rev.\ Lett.\  {\bf 21}, 1832 (1968).

\bibitem{Defoix:1969qx} 
  C.~Defoix, P.~Rivet, J.~Siaud, B.~Conforto, M.~Widgoff and F.~Shively,
  Phys.\ Lett.\  B {\bf 28}, 353 (1968).

\bibitem{Protopopescu:1973sh} 
  S.~D.~Protopopescu {\it et al.},
  Phys.\ Rev.\ D {\bf 7}, 1279 (1973).

\bibitem{Hyams:1973zf} 
  B.~Hyams {\it et al.},
  Nucl.\ Phys.\ B {\bf 64}, 134 (1973).

\bibitem{Godfrey:1985xj} 
  S.~Godfrey and N.~Isgur,
  Phys.\ Rev.\ D {\bf 32}, 189 (1985).

\bibitem{Morgan:1990kw} 
  D.~Morgan and M.~R.~Pennington,
  Z.\ Phys.\ C {\bf 48}, 623 (1990).

\bibitem{Morgan:1993td} 
  D.~Morgan and M.~R.~Pennington,
  Phys.\ Rev.\ D {\bf 48}, 1185 (1993).

\bibitem{Jaffe:1976ig} 
  R.~L.~Jaffe,
  Phys.\ Rev.\ D {\bf 15}, 267 (1977).

\bibitem{Jaffe:1976ih} 
  R.~L.~Jaffe,
  Phys.\ Rev.\ D {\bf 15}, 281 (1977).

\bibitem{Achasov:1980tb} 
  N.~N.~Achasov, S.~A.~Devyanin and G.~N.~Shestakov,
  Phys.\ Lett.\  B {\bf 96}, 168 (1980).

\bibitem{Au:1986mq} 
  K.~L.~Au, M.~R.~Pennington and D.~Morgan,
  Phys.\ Lett.\  B {\bf 167}, 229 (1986).

\bibitem{Flatte:1976xu} 
  S.~M.~Flatt\'e,
  Phys.\ Lett.\  {\bf B 63}, 224 (1976).

\bibitem{Weinstein:1982gc} 
  J.~D.~Weinstein and N.~Isgur,
  Phys.\ Rev.\ Lett.\  {\bf 48}, 659 (1982).

\bibitem{Weinstein:1990gu} 
  J.~D.~Weinstein and N.~Isgur,
  Phys.\ Rev.\ D {\bf 41}, 2236 (1990).

\bibitem{Zou:1994ea} 
  B.~S.~Zou and D.~V.~Bugg,
  Phys.\ Rev.\ D {\bf 50}, 591 (1994).

\bibitem{Janssen:1994wn} 
  G.~Janssen, B.~C.~Pearce, K.~Holinde and J.~Speth,
  Phys.\ Rev.\ D {\bf 52}, 2690 (1995)
  [nucl-th/9411021].

\bibitem{Tornqvist:1995ay} 
  N.~A.~Tornqvist and M.~Roos,
  Phys.\ Rev.\ Lett.\  {\bf 76}, 1575 (1996)
  [hep-ph/9511210].

\bibitem{Oller:1997ti} 
  J.~A.~Oller and E.~Oset,
  Nucl.\ Phys.\ A {\bf 620}, 438 (1997)
  Erratum: [Nucl.\ Phys.\ A {\bf 652}, 407 (1999)]
  [hep-ph/9702314].

\bibitem{Locher:1997gr} 
  M.~P.~Locher, V.~E.~Markushin and H.~Q.~Zheng,
  Eur.\ Phys.\ J.\ C {\bf 4}, 317 (1998)
  [hep-ph/9705230].

\bibitem{Oller:1998hw} 
  J.~A.~Oller, E.~Oset and J.~R.~Pel\'aez,
  Phys.\ Rev.\ D {\bf 59}, 074001 (1999)
  Erratum: [Phys.\ Rev.\ D {\bf 60}, 099906 (1999)]
  Erratum: [Phys.\ Rev.\ D {\bf 75}, 099903 (2007)]
  [hep-ph/9804209].

\bibitem{Oller:1998zr} 
  J.~A.~Oller and E.~Oset,
  Phys.\ Rev.\ D {\bf 60}, 074023 (1999)
  [hep-ph/9809337].

\bibitem{Baru:2003qq} 
  V.~Baru, J.~Haidenbauer, C.~Hanhart, Y.~Kalashnikova and A.~E.~Kudryavtsev,
  Phys.\ Lett.\ B {\bf 586}, 53 (2004)
  [hep-ph/0308129].

\bibitem{Baru:2004xg} 
  V.~Baru, J.~Haidenbauer, C.~Hanhart, A.~E.~Kudryavtsev and U.~G.~Mei{\ss}ner,
  Eur.\ Phys.\ J.\ A {\bf 23}, 523 (2005)
  [nucl-th/0410099].
  
\bibitem{Baru:2010ww} 
  V.~Baru, C.~Hanhart, Y.~S.~Kalashnikova, A.~E.~Kudryavtsev and A.~V.~Nefediev,
  Eur.\ Phys.\ J.\ A {\bf 44}, 93 (2010)
  [arXiv:1001.0369 [hep-ph]].

\bibitem{Bugg:1994mg} 
  D.~V.~Bugg, V.~V.~Anisovich, A.~Sarantsev and B.~S.~Zou,
  Phys.\ Rev.\ D {\bf 50}, 4412 (1994).

\bibitem{Dudek:2016cru} 
  J.~J.~Dudek {\it et al.} [Hadron Spectrum Collaboration],
  Phys.\ Rev.\ D {\bf 93}, no. 9, 094506 (2016)
  [arXiv:1602.05122 [hep-ph]].

\bibitem{Guo:2016zep} 
  Z.~H.~Guo, L.~Liu, U.~G.~Mei{\ss}ner, J.~A.~Oller and A.~Rusetsky,
  Phys.\ Rev.\ D {\bf 95}, no. 5, 054004 (2017)
  [arXiv:1609.08096 [hep-ph]].

\bibitem{GarciaMartin:2011jx} 
  R.~Garcia-Martin, R.~Kaminski, J.~R.~Pel\'aez and J.~Ruiz de Elvira,
  Phys.\ Rev.\ Lett.\  {\bf 107}, 072001 (2011)
  [arXiv:1107.1635 [hep-ph]].

\bibitem{Colangelo:2001df} 
  G.~Colangelo, J.~Gasser and H.~Leutwyler,
  Nucl.\ Phys.\ B {\bf 603}, 125 (2001)
  [hep-ph/0103088];
  B.~Ananthanarayan, G.~Colangelo, J.~Gasser and H.~Leutwyler,
  Phys.\ Rept.\  {\bf 353}, 207 (2001)
  [hep-ph/0005297].

\bibitem{Caprini:2005zr}   I.~Caprini, G.~Colangelo and H.~Leutwyler,
  Phys.\ Rev.\ Lett.\  {\bf 96}, 132001 (2006)
  [hep-ph/0512364].
  
\bibitem{pdg2018} M. Tanabashi {\it et al.} (Particle Data Group), Phys.\ Rev.\ D {\bf 98}, 030001 (2018).

\bibitem{albaladejo.190504.1} 
  M.~Albaladejo and J.~A.~Oller,
  Phys.\ Rev.\ Lett.\  {\bf 101}, 252002 (2008)
  [arXiv:0801.4929 [hep-ph]].

\bibitem{albaladejo.190504.2}
   M.~Albaladejo and J.~A.~Oller,
  Phys.\ Rev.\ D {\bf 86}, 034003 (2012)
  [arXiv:1205.6606 [hep-ph]].

\bibitem{guo.190504.1}
 Z.~H.~Guo, J.~A.~Oller and J.~Ruiz de Elvira,
  Phys.\ Lett.\ B {\bf 712}, 407 (2012)
  [arXiv:1203.4381 [hep-ph]]; 
Phys.\ Rev.\ D {\bf 86}, 054006 (2012)
  [arXiv:1206.4163 [hep-ph]].

\bibitem{GomezNicola:2001as} 
  A.~Gomez Nicola and J.~R.~Pel\'aez,
  Phys.\ Rev.\ D {\bf 65}, 054009 (2002)
  [hep-ph/0109056].

\bibitem{Guo:2011pa} 
  Z.~H.~Guo and J.~A.~Oller,
  Phys.\ Rev.\ D {\bf 84}, 034005 (2011)
  [arXiv:1104.2849 [hep-ph]].

\bibitem{Achasov:1979xc} 
  N.~N.~Achasov, S.~A.~Devyanin and G.~N.~Shestakov,
  Phys.\ Lett.\ B  {\bf 88}, 367 (1979).

\bibitem{Kerbikov:2000pu} 
  B.~Kerbikov and F.~Tabakin,
  Phys.\ Rev.\ C {\bf 62}, 064601 (2000)
  [nucl-th/0006017].

\bibitem{Close:2000ah} 
  F.~E.~Close and A.~Kirk,
  Phys.\ Lett.\ B {\bf 489}, 24 (2000)
  [hep-ph/0008066].

\bibitem{Grishina:2001zj} 
  V.~Y.~Grishina, L.~A.~Kondratyuk, M.~Buescher, W.~Cassing and H.~Stroher,
  Phys.\ Lett.\ B {\bf 521}, 217 (2001)
  [nucl-th/0103081].

\bibitem{Close:2001ay} 
  F.~E.~Close and A.~Kirk,
  Phys.\ Lett.\ B {\bf 515}, 13 (2001)
  [hep-ph/0106108].

\bibitem{Achasov:2002hg} 
  N.~N.~Achasov and A.~V.~Kiselev,
  Phys.\ Lett.\ B {\bf 534}, 83 (2002)
  [hep-ph/0203042].

\bibitem{Kudryavtsev:2002uu} 
  A.~E.~Kudryavtsev, V.~E.~Tarasov, J.~Haidenbauer, C.~Hanhart and J.~Speth,
  Phys.\ Rev.\ C {\bf 66}, 015207 (2002)
  [nucl-th/0203034].

\bibitem{Achasov:2003se} 
  N.~N.~Achasov and G.~N.~Shestakov,
  Phys.\ Rev.\ Lett.\  {\bf 92}, 182001 (2004)
  [hep-ph/0312214].

\bibitem{Achasov:2004ur} 
  N.~N.~Achasov and G.~N.~Shestakov,
  Phys.\ Rev.\ D {\bf 70}, 074015 (2004)
  [hep-ph/0405129].

\bibitem{Wu:2007jh} 
  J.~J.~Wu, Q.~Zhao and B.~S.~Zou,
  Phys.\ Rev.\ D {\bf 75}, 114012 (2007)
  [arXiv:0704.3652 [hep-ph]].

\bibitem{Hanhart:2007bd} 
  C.~Hanhart, B.~Kubis and J.~R.~Pel\'aez,
  Phys.\ Rev.\ D {\bf 76}, 074028 (2007)
  [arXiv:0707.0262 [hep-ph]].

\bibitem{Wu:2008hx} 
  J.~J.~Wu and B.~S.~Zou,
  Phys.\ Rev.\ D {\bf 78}, 074017 (2008)
  [arXiv:0808.2683 [hep-ph]].

\bibitem{Ablikim:2010aa} 
  M.~Ablikim {\it et al.} [BESIII Collaboration],
  Phys.\ Rev.\ D {\bf 83}, 032003 (2011)
  [arXiv:1012.5131 [hep-ex]].

\bibitem{Ablikim:2018pik} 
  M.~Ablikim {\it et al.} [BESIII Collaboration],
  Phys.\ Rev.\ Lett.\  {\bf 121}, no. 2, 022001 (2018)
  [arXiv:1802.00583 [hep-ex]].

\bibitem{Oset:1997it} 
  E.~Oset and A.~Ramos,
  Nucl.\ Phys.\ A {\bf 635}, 99 (1998)
  [nucl-th/9711022].

\bibitem{Oller:2000fj} 
  J.~A.~Oller and U.~G.~Mei{\ss}ner,
  Phys.\ Lett.\ B {\bf 500}, 263 (2001)
  [hep-ph/0011146].

\bibitem{Roca:2012cv} 
  L.~Roca,
  Phys.\ Rev.\ D {\bf 88}, 014045 (2013)
  [arXiv:1210.4742 [hep-ph], arXiv:1210.4742 [hep-ph]].

\bibitem{Bayar:2017pzq} 
  M.~Bayar and V.~R.~Debastiani,
  Phys.\ Lett.\ B {\bf 775}, 94 (2017)
  [arXiv:1708.02764 [hep-ph]].

\bibitem{Sekihara:2014qxa} 
  T.~Sekihara and S.~Kumano,
  Phys.\ Rev.\ D {\bf 92}, no. 3, 034010 (2015)
  [arXiv:1409.2213 [hep-ph]].

\bibitem{Nussinov:1989gs} 
  S.~Nussinov and T.~N.~Truong,
  Phys.\ Rev.\ Lett.\  {\bf 63}, 1349 (1989)
  Erratum: [Phys.\ Rev.\ Lett.\  {\bf 63}, 2002 (1989)].

\bibitem{Achasov:1987ts} 
  N.~N.~Achasov and V.~N.~Ivanchenko,
  Nucl.\ Phys.\ B {\bf 315}, 465 (1989).

\bibitem{LucioMartinez:1990uw} 
  J.~L.~Lucio Martinez and J.~Pestieau,
  Phys.\ Rev.\ D {\bf 42}, 3253 (1990).

\bibitem{Close:1992ay} 
  F.~E.~Close, N.~Isgur and S.~Kumano,
  Nucl.\ Phys.\ B {\bf 389}, 513 (1993)
  [hep-ph/9301253].

\bibitem{Bramon:2000vu} 
  A.~Bramon, R.~Escribano, J.~L.~Lucio M., M.~Napsuciale and G.~Pancheri,
  Phys.\ Lett.\ B {\bf 494}, 221 (2000)
  [hep-ph/0008188].

\bibitem{Bramon:2001un} 
  A.~Bramon, R.~Escribano, J.~L.~Lucio Martinez and M.~Napsuciale,
  Phys.\ Lett.\ B {\bf 517}, 345 (2001)
  [hep-ph/0105179].

\bibitem{Bramon:2002iw} 
  A.~Bramon, R.~Escribano, J.~L.~Lucio M, M.~Napsuciale and G.~Pancheri,
  Eur.\ Phys.\ J.\ C {\bf 26}, 253 (2002)
  [hep-ph/0204339].

\bibitem{Oller:1998ia} 
  J.~A.~Oller,
  Phys.\ Lett.\ B {\bf 426}, 7 (1998)
  [hep-ph/9803214].

\bibitem{Marco:1999df} 
  E.~Marco, S.~Hirenzaki, E.~Oset and H.~Toki,
  Phys.\ Lett.\ B {\bf 470}, 20 (1999)
  [hep-ph/9903217].

\bibitem{Palomar:2003rb} 
  J.~E.~Palomar, L.~Roca, E.~Oset and M.~J.~Vicente Vacas,
  Nucl.\ Phys.\ A {\bf 729}, 743 (2003)
  [hep-ph/0306249].

\bibitem{Ablikim:2016exh} 
  M.~Ablikim {\it et al.} [BESIII Collaboration],
  Phys.\ Rev.\ D {\bf 94}, no. 7, 072005 (2016)
  [arXiv:1608.07393 [hep-ex]].


 \bibitem{Pelaez:2006nj} 
  J.~R.~Pel\'aez and G.~Rios,
  Phys.\ Rev.\ Lett.\  {\bf 97}, 242002 (2006)
  doi:10.1103/PhysRevLett.97.242002
  [hep-ph/0610397].

\bibitem{Pelaez:2010fj} 
  J.~R.~Pel\'aez and G.~R\'{\i}os,
  Phys.\ Rev.\ D {\bf 82}, 114002 (2010)
  doi:10.1103/PhysRevD.82.114002
  [arXiv:1010.6008 [hep-ph]].

\bibitem{Aceti:2015zva} 
  F.~Aceti, J.~M.~Dias and E.~Oset,
  Eur.\ Phys.\ J.\ A {\bf 51}, no. 4, 48 (2015)
  [arXiv:1501.06505 [hep-ph]].




\bibitem{hp.190506.2}  T.~A.~Armstrong {\it et al.} [WA76 and Athens-Bari-Birmingham-CERN-College de France Collaborations],
  Z.\ Phys.\ C {\bf 52}, 389 (1991).

\bibitem{hp.190506.1}
 J.~B.~Gay {\it et al.} [Amsterdam-CERN-Nijmegen-Oxford Collaboration],
  Phys.\ Lett.\  {\bf 63B}, 220 (1976).

\bibitem{Oller:2002na} 
  J.~A.~Oller,
  Nucl.\ Phys.\ A {\bf 714}, 161 (2003)
  [hep-ph/0205121].

\bibitem{Bramon:1992ki} 
  A.~Bramon, A.~Grau and G.~Pancheri,
  Phys.\ Lett.\ B {\bf 289}, 97 (1992).

\bibitem{Meissner:1987ge} 
  U.~G.~Mei{\ss}ner,
  Phys.\ Rept.\  {\bf 161}, 213 (1988).
  
\bibitem{Xiao:2012iq} 
  C.~W.~Xiao and E.~Oset,
  Eur.\ Phys.\ J.\ A {\bf 49}, 52 (2013)
  [arXiv:1211.1862 [hep-ph]].

\bibitem{Klingl:1996by} 
  F.~Klingl, N.~Kaiser and W.~Weise,
  Z.\ Phys.\ A {\bf 356}, 193 (1996)
  [hep-ph/9607431].

\bibitem{Meissner:2000bc} 
  U.~G.~Mei{\ss}ner and J.~A.~Oller,
  Nucl.\ Phys.\ A {\bf 679}, 671 (2001)
  [hep-ph/0005253].

\bibitem{Micu:1968mk} 
  L.~Micu,
  Nucl.\ Phys.\ B {\bf 10}, 521 (1969).

\bibitem{Ablikim:2015umt} 
  M.~Ablikim {\it et al.} [BESIII Collaboration],
  Phys.\ Rev.\ D {\bf 92}, no. 5, 052003 (2015)
  Erratum: [Phys.\ Rev.\ D {\bf 93}, no. 3, 039906 (2016)]
  [arXiv:1506.00546 [hep-ex]].

\bibitem{Ablikim:2006db} 
  M.~Ablikim {\it et al.},
  Phys.\ Lett.\ B {\bf 642}, 441 (2006)
  [hep-ex/0603048].

\bibitem{Becker:1986zt} 
  J.~Becker {\it et al.} [Mark-III Collaboration],
  Phys.\ Rev.\ D {\bf 35}, 2077 (1987).

\bibitem{Augustin:1987da} 
  J.~E.~Augustin {\it et al.} [DM2 Collaboration],
  Z.\ Phys.\ C {\bf 36}, 369 (1987).

\bibitem{Bai:1996wm} 
  J.~Z.~Bai {\it et al.} [BES Collaboration],
  Phys.\ Rev.\ Lett.\  {\bf 76}, 3502 (1996).

\bibitem{Ablikim:2018izx} 
  M.~Ablikim {\it et al.} [BESIII Collaboration],
  arXiv:1808.06946 [hep-ex].

\bibitem{MartinezTorres:2012du} 
  A.~Martinez Torres, K.~P.~Khemchandani, F.~S.~Navarra, M.~Nielsen and E.~Oset,
  Phys.\ Lett.\ B {\bf 719}, 388 (2013)
  [arXiv:1210.6392 [hep-ph]].

\bibitem{Xie:2015lta} 
  J.~J.~Xie and E.~Oset,
  Phys.\ Lett.\ B {\bf 753}, 591 (2016)
  [arXiv:1509.08099 [hep-ph]].



\end{thebibliography}
\end{document}